\documentclass[aps, pra, twocolumn, superscriptaddress, amsmath,  tightenlines, longbibliography]{revtex4-1}

\usepackage{dcolumn}
\usepackage{lipsum} 
\usepackage{graphicx}
\usepackage{mathrsfs}
\usepackage{subfigure}
\usepackage{booktabs}
\usepackage{amsmath}
\usepackage{physics}
\usepackage{appendix}
\usepackage{dsfont}
\usepackage{amstext}
\usepackage{amssymb}
\usepackage{amsbsy}
\usepackage{bbm}
\usepackage{amsthm}
\usepackage{graphicx}
\usepackage{color}
\usepackage[colorlinks,citecolor=blue]{hyperref}

\usepackage{url}
\usepackage[colorlinks]{hyperref}
\hypersetup{%
	plainpages=true,
	breaklinks=true,       
	hypertexnames=false,  
	pageanchor=true,
	colorlinks=true,
	linkcolor={blue},
	citecolor={red},
	urlcolor={blue},
	anchorcolor={black}
}

 \makeatletter

\newcommand{\Rmnum}[1]{\expandafter\@slowromancap\romannumeral #1@}
\makeatother

\begin{document}
\title{Non-Hermitian Topological Mott Insulators in 1D Fermionic Superlattices}
\author{Tao Liu}
\email[E-mail: ]{tao.liu@riken.jp}
\affiliation{Theoretical Quantum Physics Laboratory, RIKEN Cluster for Pioneering Research, Wako-shi, Saitama 351-0198, Japan}
\affiliation{School of Physics and Optoelectronics, South China University of Technology, Guangzhou 510640, China}

\author{James Jun He}
\affiliation{RIKEN Center for Emergent Matter Science, Wako, Saitama 351-0198, Japan}

\author{Tsuneya Yoshida}
\affiliation{Graduate School of Pure and Applied Sciences, University of Tsukuba, Tsukuba, Ibaraki, 305-8571, Japan.}
\affiliation{Department of Physics, University of Tsukuba, Tsukuba, Ibaraki, 305-8571, Japan.}

\author{Ze-Liang Xiang}
\affiliation{School of Physics, Sun Yat-sen University, Guangzhou 510275, China}

\author{Franco Nori}
\email[E-mail: ]{fnori@riken.jp}
\affiliation{Theoretical Quantum Physics Laboratory, RIKEN Cluster for Pioneering Research, Wako-shi, Saitama 351-0198, Japan}
\affiliation{Department of Physics, University of Michigan, Ann Arbor, Michigan 48109-1040, USA}

\date{{\small \today}}


\begin{abstract}
	We study interaction-induced Mott insulators, and their topological properties in a 1D non-Hermitian strongly-correlated spinful fermionic superlattice system with either nonreciprocal hopping or complex-valued interaction. For the nonreciprocal hopping case, the low-energy neutral excitation spectrum is sensitive to boundary conditions, which is a manifestation of the non-Hermitian skin effect. However, unlike the single-particle case, particle density of strongly correlated system does not suffer from the non-Hermitian skin effect due to the Pauli exclusion principle and repulsive interactions. Moreover, the anomalous boundary effect occurs due to the interplay of nonreciprocal hopping, superlattice potential, and strong correlations,  where some in-gap modes, for both the neutral and charge excitation spectra, show no edge excitations defined via only the right eigenvectors. We show that these edge excitations of the in-gap states can be correctly characterized by only biorthogonal eigenvectors. Furthermore, the topological Mott phase, with gapless particle excitations around  boundaries, exists even for the purely imaginary-valued interaction, where the continuous quantum Zeno effect leads to the effective on-site repulsion between two-component fermions.   
\end{abstract}

\maketitle

\section{Introduction} 
Recent years have witnessed considerable interest in exploring topological phases of non-Hermitian systems \cite{PhysRevLett.102.065703, PhysRevB.84.205128, PhysRevLett.115.200402, Weimann2016, PhysRevLett.116.133903, PhysRevLett.118.045701, PhysRevLett.118.040401, PhysRevLett.120.146402, PhysRevB.97.121401, Harari2018, Bandreseaar4005, Zhoueaap9859, Xiong2018, Pan2018, arXiv:1802.07964, arXiv:1806.06566,  arXiv:1805.06492,  ShunyuYao2018, YaoarXiv:1804.04672, KoheiarXiv:1805.09632, PhysRevB.100.054105, Kawabata2019, PhysRevB.99.201103,  PhysRevB.99.235112, PhysRevX.9.041015, PhysRevB.100.144106, PhysRevB.100.035102, PhysRevLett.122.076801, arXiv:1905.02211, PhysRevLett.123.170401, Ghatak_2019, PhysRevB.99.121101, PhysRevLett.123.066405, PhysRevLett.123.097701, arXiv:1910.02878, arXiv:1901.08060, PhysRevLett.123.016805, PhysRevB.99.081302, PhysRevLett.123.066404, PhysRevLett.123.073601, PhysRevResearch.1.023013, PhysRevB.99.201411, PhysRevB.100.184314, arXiv:1905.07109, arXiv:1904.08724, arXiv:1904.02492, PhysRevB.100.045141, arXiv:1906.03988, PhysRevLett.123.206404, PhysRevB.100.155117, PhysRevLett.123.206404, arXiv:1910.10946, Zhao2019, arXiv:1912.12022, arXiv:1912.04024, arXiv:1901.11241, arXiv:1902.07217, arXiv:1912.05825, arXiv:1912.10048, arXiv:1911.01590}, which can be realized in classical optical and mechanical systems with gain and loss \cite{PhysRevLett.100.103904, PhysRevLett.106.093902, Regensburger2012, PhysRevLett.113.053604, Hodaei975, Peng2014, Feng972, Peng328, Jing2015, PhysRevLett.117.110802, PhysRevLett.119.190401, Jing2017, PhysRevApplied.8.044020,  El-Ganainy2018, Zhang2018, arXiv:2004.09529}, correlated and disordered electronic systems with finite quasiparticle lifetimes \cite{arXiv:1708.05841, arXiv:1802.03023, PhysRevB.97.041203, PhysRevB.99.201107, PhysRevB.98.035141}, and open quantum systems with post-selection measurements \cite{Naghiloo2019}.  Non-Hermitian topological systems exhibit many unique properties with no counterpart in Hermitian cases, such as: non-Hermitian skin effect and breakdown of the conventional bulk-boundary correspondence \cite{arXiv:1805.06492, ShunyuYao2018, YaoarXiv:1804.04672}, bulk Fermi arcs \cite{Zhoueaap9859}, and rich Bernard-LeClair symmetry classes beyond the ten-fold Altland-Zirnbauer ones \cite{PhysRevB.99.235112, PhysRevX.9.041015, PhysRevB.100.144106}. However, most of these studies have focused on non-Hermitian topological band theories at the single-particle level, and only very few recent works \cite{arXiv:1911.01590,Yoshida2019} have studied strongly correlated non-Hermitian topological phases.

Rather than destroy the topological properties, the strong correlation can give rise to novel topological phases for Hermitian systems \cite{Sheng2011, PhysRevX.1.021014,  BERGHOLTZ2013,  Maciejko2015, PhysRevLett.121.025301, PhysRevLett.100.156401, Rachel2018, Pesin2010,  PhysRevLett.112.196404,   PhysRevLett.123.196402}. For example, interactions lead to fractional Chern insulators  with exotic fractional quasiparticles obeying fractional statistics \cite{ Sheng2011, PhysRevX.1.021014, BERGHOLTZ2013,  Maciejko2015}. In addition, interactions result in topological Mott insulators \cite{PhysRevLett.100.156401, Rachel2018}: interactions open a nontrivial gap in the bulk, inducing single-particle gapless excitations around the boundary \cite{NoteMottNonHermitian}. Up to now, studies of strongly correlated topological phases have been largely restricted to Hermitian systems. It is important to explore how topology and interactions interplay with non-Hermitian effects, leading to novel topological features without its Hermitian counterpart. In particular, in the single-particle case, nonreciprocal hopping causes the localization of eigenstates to the boundaries. This non-Hermitian skin effect, which leads to the breakdown of the bulk-boundary correspondence, does not cause the same localization in a fermionic many-body system \cite{arXiv:1912.05825}. This leads us to ask: Does nonreciprocal hopping lead to anomalous boundary effects in a  non-Hermitian strongly correlated topological system? In addition to the nonreciprocal hopping case, can the purely imaginary-valued interaction drive a topologically trivial system into topologically nontrivial regime?

In this paper, we address these important questions by studying a non-Hermitian strongly correlated system in a 1D spin-1/2 fermionic superlattice system. The non-Hermitian Hamiltonian is constructed by introducing either nonreciprocal hopping or complex-valued interactions. For both cases, the interaction can drive the trivial non-Hermitian system into the topological Mott phase, and there exist both gapless neutral and charge excitations localized at the edges. For the nonreciprocal hopping case, the low-energy neutral excitation spectrum is sensitive to the boundary conditions, where the in-gap states emerge in the open chain in spite of the absence of gap in the periodic chain. This shows the impact of the non-Hermitian skin effect. However, unlike the single-particle case, the non-Hermitian skin effect plays no role in particle density due to the Pauli exclusion principle and repulsive interactions. We also found that some in-gap states, for both neutral and charge excitation spectra, show no edge excitations defined via only the right eigenvectors. This anomalous boundary effect results from the interplay of nonreciprocal hopping, superlattice potential, and interactions. We show that these edge excitations can be correctly characterized by using biorthogonal eigenvectors. Furthermore, the purely imaginary-valued interaction can also drive the trivial system to topological Mott insulator, where the effective repulsion is created by the continuous quantum Zeno effect.

The rest of this paper is organized as follows. In Sec.~$\textrm{\Rmnum{2}}$, we consider the non-Hermitian Fermi-Hubbard model in the presence of superlattice modulation and nonreciprocal hopping, and discuss the edge neutral and charge excitations. In Sec.~$\textrm{\Rmnum{3}}$, we investigate the topological Mott phase driven by complex-valued interactions. We conclude this work in Sec. $\textrm{\Rmnum{4}}$.   

\begin{figure}[!b]
	\centering
	\includegraphics[width=8.4cm]{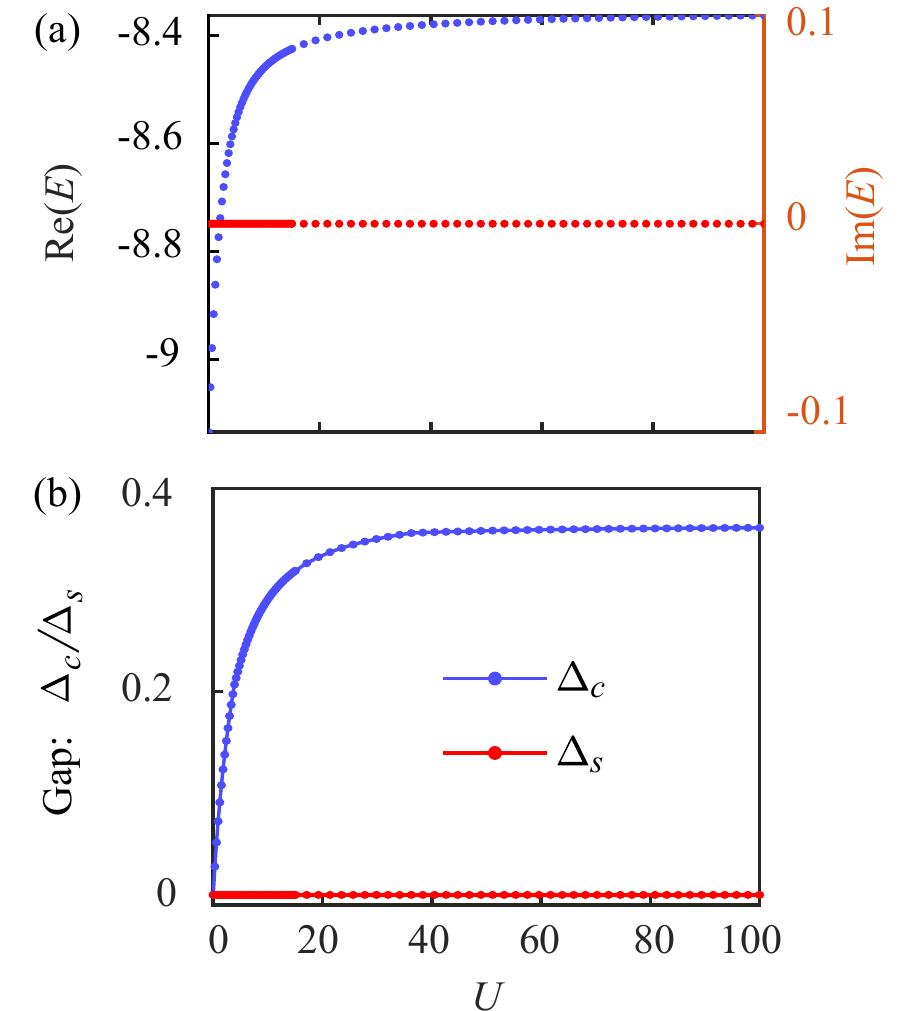}
	\caption{(a) Real (blue dots) and imaginary (red dots) parts of ground states versus the interaction strength for the periodic boundary condition. As the on-site energy increases, the real parts of the eigenenergies first rises, and then tend to a finite value.  (b) Charge gap $\Delta_c$ and spin gap $\Delta_s$ versus the interaction strength $U$ for periodic boundary. A non-zero charge gap indicates the Mott phase driven by the interactions. The parameters used are: $L = 12$, $N_\uparrow = 2$, $N_\downarrow = 2$, $t=1$, $V_0=1.5$, $q = 3$, $\phi = 2 \pi/3$, and $\alpha = 0.4$.  }\label{fig1}
\end{figure}
\begin{figure*}[!tb]
	\centering
	\includegraphics[width=15cm]{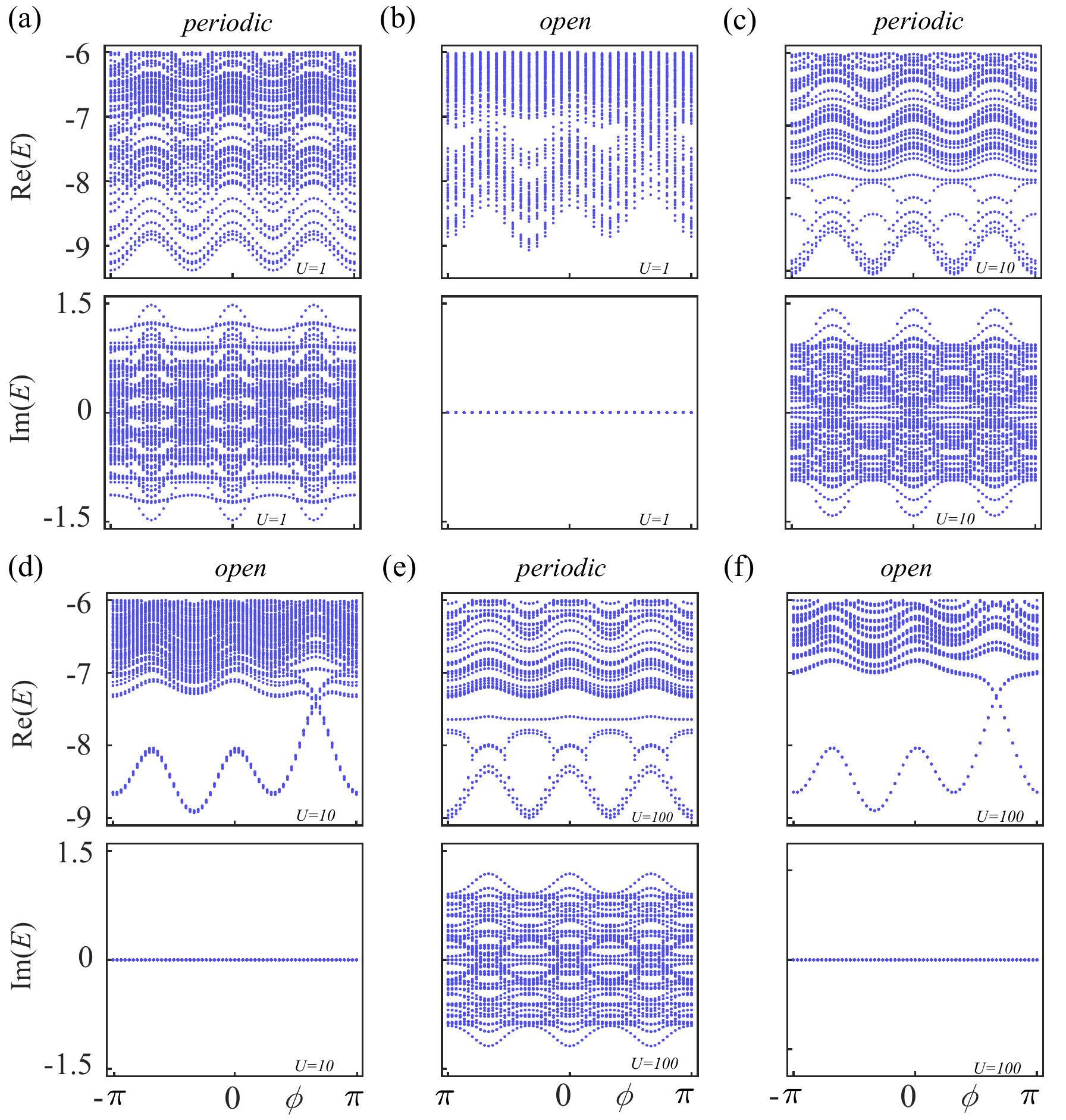}
	\caption{Low-energy spectra versus the modulation phase $\phi$ for: (a,b)  $U = 1$, (c,d)  $U = 10$, and (e,f) $U = 100$. The spectra in (a,c,e) are for periodic boundary, and (b,d,f) for open boundaries. In  the stronge interaction limit, gapless neutral excitations emerge for open boundaries, while no gap is opened for periodic boundaries. This indicates that the spectrum of the non-Hermitian correlated system with nonreciprocal hopping is sensitive to the boundary conditions, which is manifestation of the non-Hermitian skin effect.  The parameters used here are: $L = 12$, $N_\uparrow = 2$, $N_\downarrow = 2$, $t=1$, $V_0=1.5$, $q = 3$, and $\alpha = 0.4$. }\label{fig2}
\end{figure*}
\begin{figure}[!tb]
	\centering
	\includegraphics[width=8.4cm]{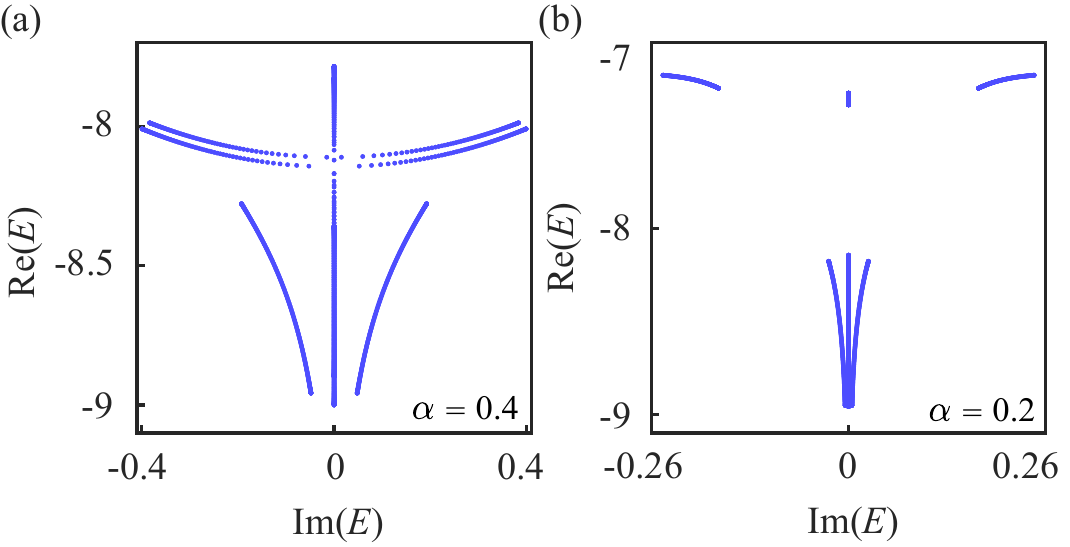}
	\caption{Few lowest eigenenergies in the complex plane using periodic boundary, for (a) $\alpha=0.4$, and (b) $\alpha=0.2$.  The parameters used here are: $L = 12$, $N_\uparrow = 2$, $N_\downarrow = 2$, $t=1$, $V_0=1.5$, $q = 3$, and $U = 100$. For the weak nonreciprocal hopping (i.e., $\alpha=0.2$), the many-body spectrum shows a band gap, while, in the strong nonreciprocal hopping (i.e., $\alpha=0.4$), no gap is opened for periodic boundaries.}\label{FigN1}
\end{figure}
\begin{figure}[!tb]
	\centering
	\includegraphics[width=8.4cm]{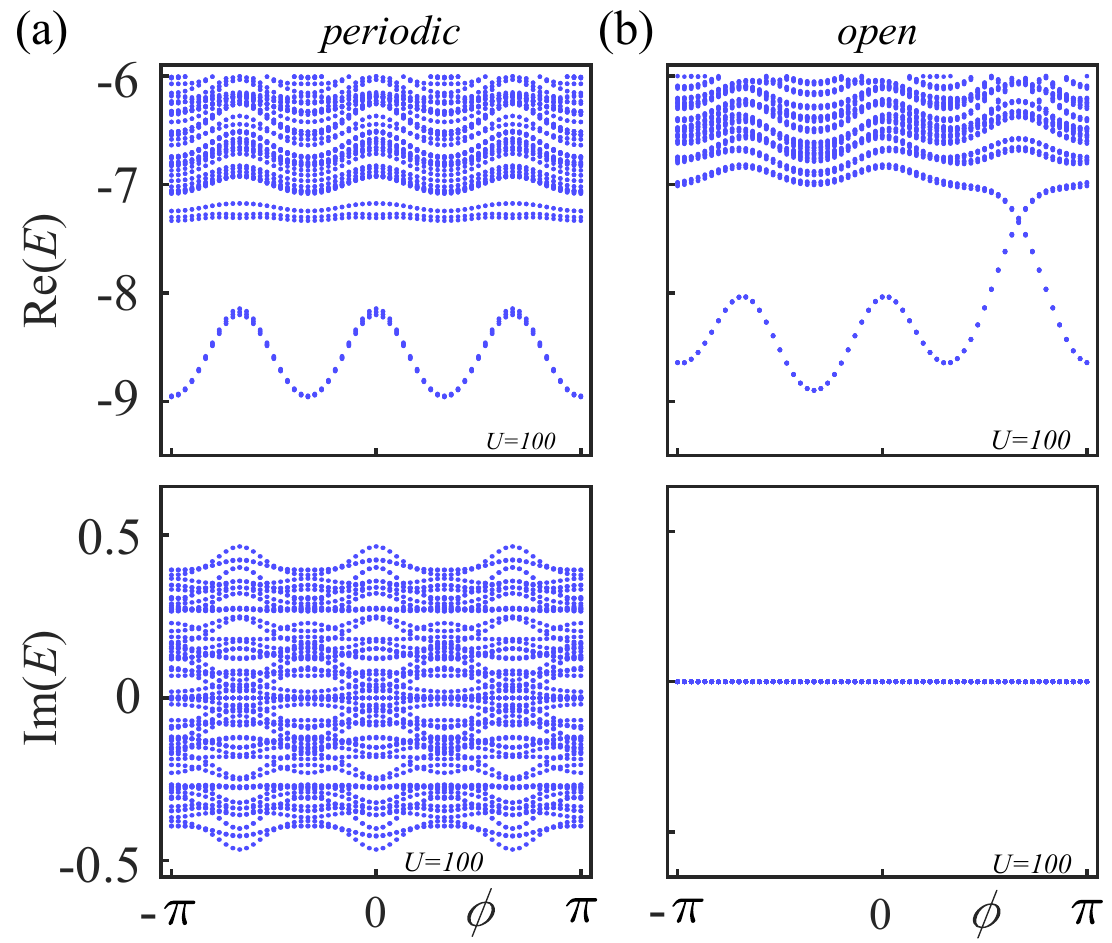}
	\caption{Low-energy spectra versus the modulation phase $\phi$ for $\alpha = 0.2$, and $U = 100$. The spectrum of (a) is for periodic boundary, and (b) for open boundaries. In contrast to the case of the strong nonreciprocal hopping, the energy spectra with weak nonreciprocal hopping are gaped.  The parameters used here are: $L = 12$, $N_\uparrow = 2$, $N_\downarrow = 2$, $t=1$, $V_0=1.5$, and $q = 3$. }\label{figNS2}
\end{figure}
\begin{figure*}[!tb]
	\centering
	\includegraphics[width=17cm]{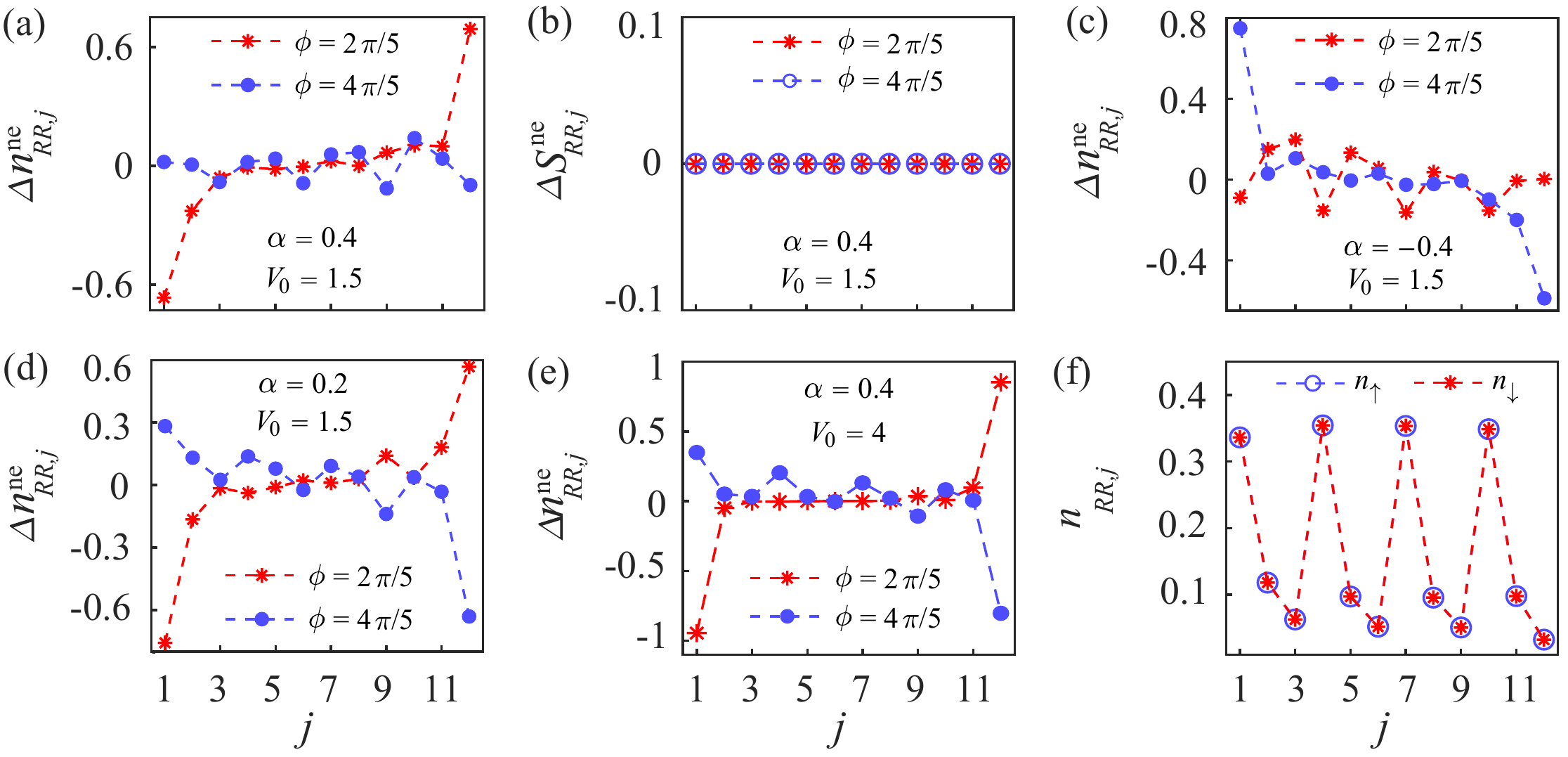}
	\caption{Spatial distributions of the (a,c,d,e) charge and (b) spin for the neutral excitation modes, calculated via only the right eigenvectors, for different modulation phases $\phi$. The spatial distributions  are calculated for: (a,b) $\alpha = 0.4$, $V_0 = 1.5$; (c) $\alpha = -0.4$, $V_0 = 1.5$; (d) $\alpha = 0.2$, $V_0 = 1.5$; and (e) $\alpha = 0.4$, $V_0 = 4$. (f) Spin-resolved charge densities of the ground state for $\phi = 2\pi/5$, $\alpha = 0.4$ and $V_0 = 1.5$. The parameters used here are: $L = 12$, $N_\uparrow = 2$, $N_\downarrow = 2$, $t=1$, $U = 100$, and $q = 3$.}\label{fig3}
\end{figure*}

\begin{figure*}[!tb]
	\centering
	\includegraphics[width=18cm]{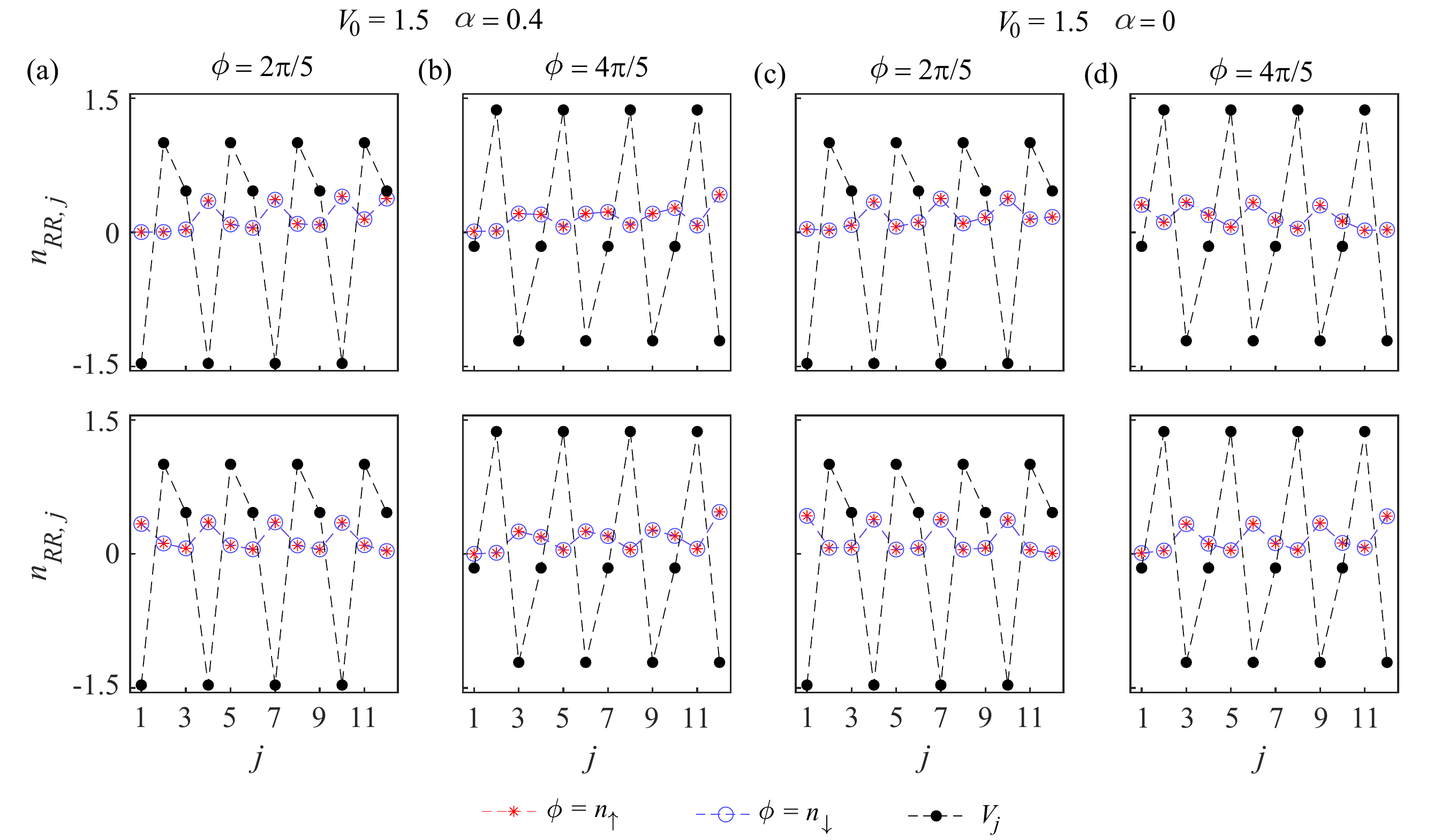}
	\caption{ Spin-resolved charge densities, calculated via only the right eigenvectors, of the ground state $\ket{\Phi_0 \left(N_\uparrow, N_\downarrow\right)}_R$ (the second row) and the higher energy state $\ket{\Phi_1 \left(N_\uparrow, N_\downarrow\right)}_R$ (the first row)  for (a,b) $\alpha=0.4$, and (c,d) $\alpha=0$ using the open boundaries. The black dotted line indicates the superlattice potential $V_j = V_0 \cos(2 \pi j/q  + \phi)$. The parameters used here are: $L = 12$, $N_\uparrow = 2$, $N_\downarrow = 2$, $t=1$, $q = 3$, $V_0=1.5$, and $U = 100$.}\label{figNS4}
\end{figure*}

\begin{figure}[!tb]
	\centering
	\includegraphics[width=7.4cm]{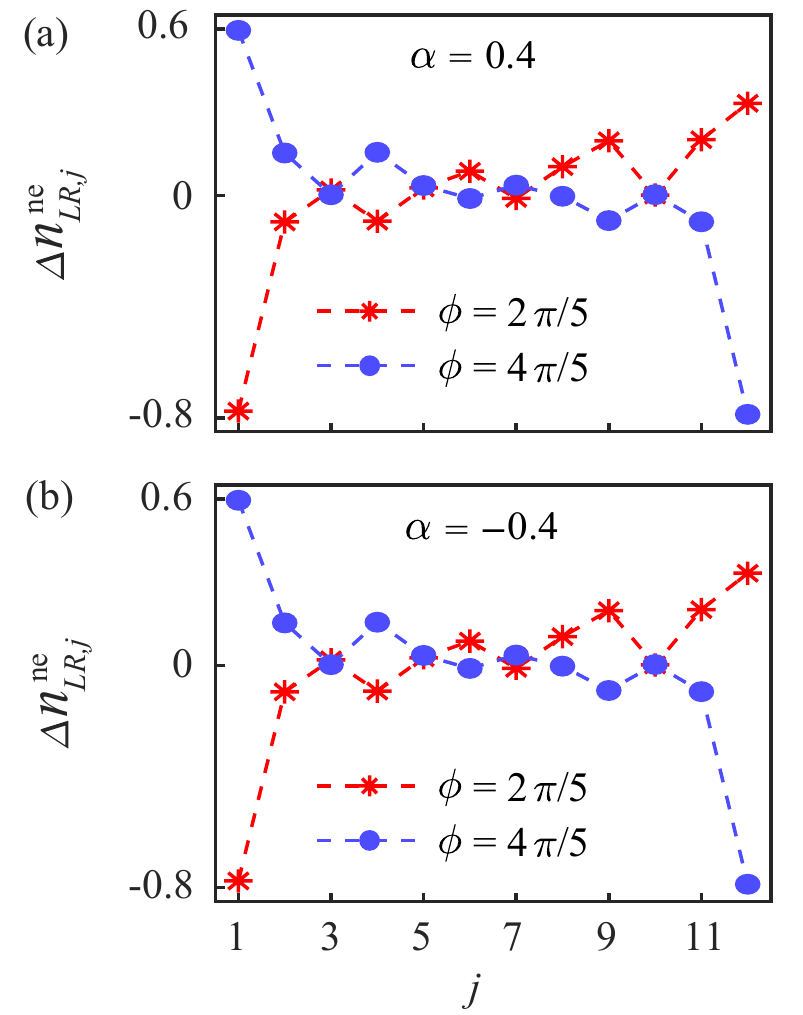}
	\caption{Spatial charge distributions, calculated via biorthogonal eigenvectors, of the neutral excitations  for (a) $\alpha=0.4$ and (b) $\alpha=-0.4$. The absence of edge excitations calculated by only the right eigenvectors is restored by the biorthogonal formula. The parameters used here are: $L = 12$, $N_\uparrow = 2$, $N_\downarrow = 2$, $t=1$, $V_0=1.5$, and $q = 3$.  }\label{fig32}
\end{figure}
\begin{figure*}[!tb]
	\centering
	\includegraphics[width=18.0cm]{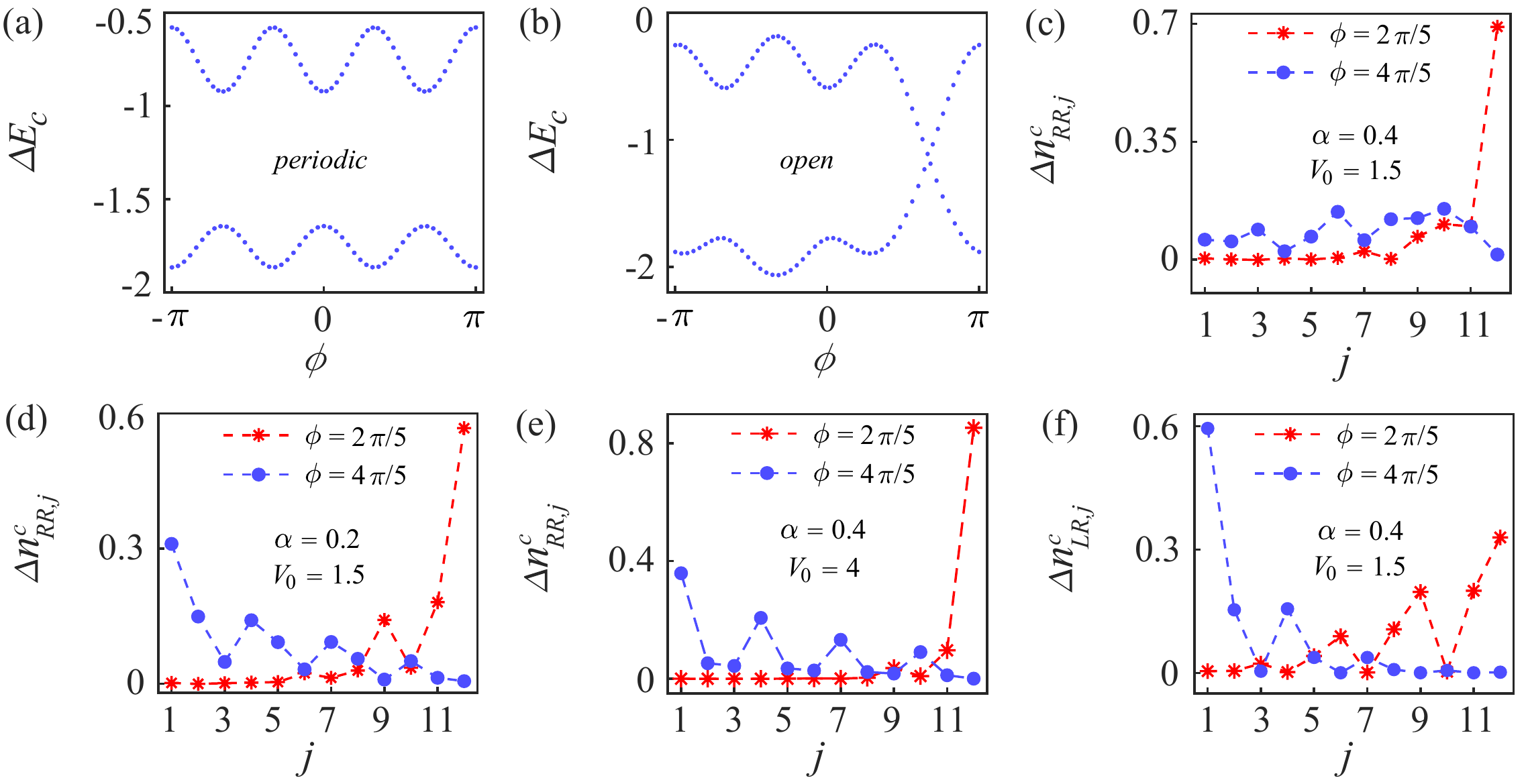}
	\caption{Charge excitation spectra $\Delta E_c(2, 2)$ and $\Delta E_c(2, 1)$ versus the modulation phase $\phi$ for (a) periodic and (b) open boundaries.  Spatial charge distributions, calculated using only the right eigevectors, of the charge excitations for: (c) $\alpha = 0.4$, $V_0 = 1.5$; (d) $\alpha = 0.2$, $V_0 = 1.5$; and (e) $\alpha = 0.4$, $V_0 = 4$. (f) Spatial charge distributions calculated using biorthogonal eigevectors. The parameters used here are: $L = 12$, $N_\uparrow = 2$, $N_\downarrow = 2$, $t=1$, and $q = 3$}\label{fig4}
\end{figure*}

\begin{figure*}[!tb]
	\centering
	\includegraphics[width=18.cm]{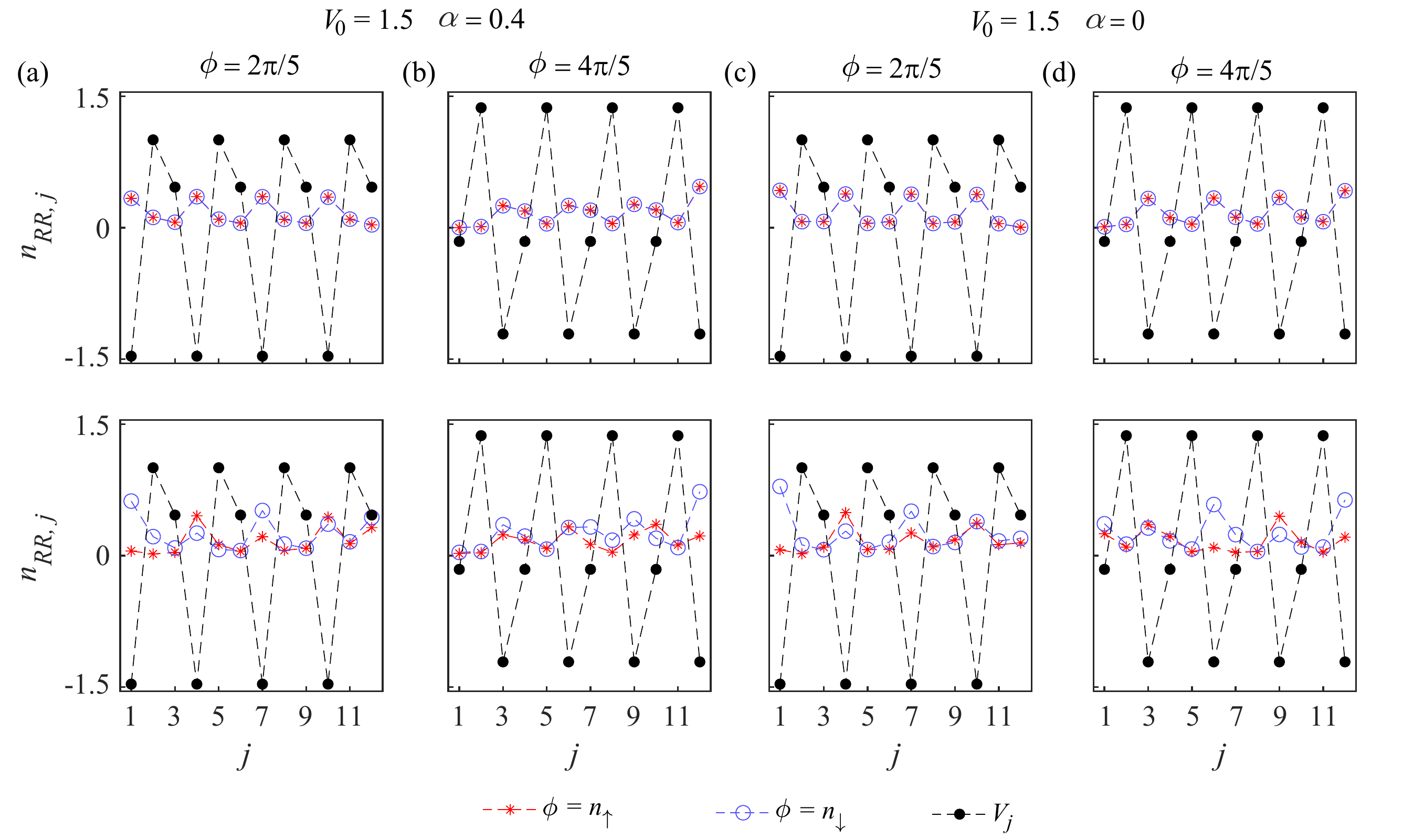}
	\caption{ Spin-resolved charge densities, calculated via only the right eigenvectors, of the ground states $\ket{\Phi_0 \left(N_\uparrow, N_\downarrow\right)}_R$ (the  first row) and $\ket{\Phi_0 \left(N_\uparrow, N_\downarrow+1\right)}_R$ (the second row) for (a,b) $\alpha=0.4$, and (c,d) $\alpha=0$ using the open boundaries. The black dotted line indicates the superlattice potential $V_j = V_0 \cos(2 \pi j/q  + \phi)$. The parameters used here are: $L = 12$, $N_\uparrow = 2$, $N_\downarrow = 2$, $t=1$, $q = 3$, $V_0=1.5$, and $U = 100$.}\label{figNS6}
\end{figure*}

\section{Non-Hermitian Model with nonreciprocal hopping}
\subsection{Model}
We consider a 1D model of spin-1/2 ultracold fermionic atoms loaded in a bichromatic optical lattice (see Appendix \ref{Appendix_A} for details), described by 
\begin{align}\label{p1_H11}
	\mathcal{H} = & -t \sum_{j,\sigma} \left( e^{\alpha} c_{j+1,\sigma}^\dagger c_{j,\sigma} + e^{-\alpha} c_{j,\sigma}^\dagger c_{j+1,\sigma}\right) + \sum_{j,\sigma} V_j n_{j,\sigma} \nonumber \\
	& + U \sum_{j} c_{j,\uparrow}^\dagger c_{j,\downarrow}^\dagger c_{j,\downarrow} c_{j,\uparrow},  
\end{align}
where $V_j = V_0 \cos(2 \pi j/q  + \phi)$ denotes a commensurate superlattice potential with the modulation period determined by $q$ and phase $\phi$, $c_{j,\sigma}^\dagger$ is the fermionic creation operator with spin $\sigma$, $n_{j,\sigma} = c_{j,\sigma}^\dagger c_{j,\sigma}$ is the on-site particle number operator, $U$ denotes the on-site interaction strength, and $\alpha$ induces nonreciprocal hopping.

In the absence of on-site interactions (see Appendix \ref{Appendix_B}), the single-particle energy spectrum of a 1D superlattice is split into $q$ subbands, and the insulator with the fully filled subband exhibits a topological phase characterized by the Chern number when $\phi$ is taken as an additional dimension \cite{PhysRevLett.108.220401}. Moreover, the Hermitian superlattice system supports topological Mott phases under the strong interaction for both fermions and bosons \cite{PhysRevLett.110.075303, PhysRevB.88.045110, PhysRevLett.110.260405, PhysRevA.100.023616}.  In this paper, we discuss the strongly correlated topological phases in the non-Hermitian case.

We consider a spin-1/2 fermionic chain of  $L$ sites, with fractional filling factors $v_\sigma = N_\sigma/N_{\textrm{cell}} = 1/2$. Here $N_\sigma$ is the number of fermions with spin $\sigma$, and $N_{\textrm{cell}} = L/q$ is the number of primitive cells. Due to half-filling at the lowest subband, the system is topologically trivial at the single-particle level. When the on-site interaction is introduced, we compute the real and imaginary parts of the eigenenergies of ground states as a function of the on-site interaction strength for filling factors $v_\sigma = 1/2$, as shown in Fig.~\ref{fig1}(a). Here, as a natural extension of the Hermitian systems, the ground state of the many-body spectrum is defined as the state with minimum real part of the eigenenergies. According to Fig.~\ref{fig1}(a), the real part of the eigenenergies of the many-body spectrum first increases, and then tends to a finite value, as the on-site interaction strength rises.

In the presence of the on-site interactions, we calculate both the charge gap $\Delta_c$ and spin gap $\Delta_s$, which are defined as 
\begin{align}\label{p1_H111}
\Delta_c = &\frac{1}{2}\left[E_0\left(N_\uparrow, N_\downarrow + 1\right) + E_0\left(N_\uparrow, N_\downarrow - 1\right)\right]  \nonumber \\ & -   E_0\left(N_\uparrow, N_\downarrow \right),  
\end{align}
\begin{align}\label{p1_H112}
\Delta_s = &\frac{1}{2}\left[E_0\left(N_\uparrow - 1, N_\downarrow + 1\right) + E_0\left(N_\uparrow + 1, N_\downarrow - 1\right) \right]  \nonumber \\ & -  2 E_0\left(N_\uparrow, N_\downarrow \right)  
\end{align}
where $E_0$ is the ground-state energy, defined as the minimum real part of the many-body energy spectrum \cite{PhysRevA.94.053615, Yoshida2019}. 

Figure \ref{fig1}(b) plots the charge and spin gaps versus the on-site interaction strength $U$. In the noninteracting limit, the lowest subband is only partially filled for $v_\sigma = 1/2$; therefore, the charge gap is zero. Once the interaction is introduced, the repulsion between the two spin components forces the fermions with different spins to occupy the individual energy level of the lowest subband. Consequently, a charge gap is opened, and the non-Hermitian metallic system becomes a Mott insulator. In contrast, the spin modes are always gapless. 

Moreover, the charge gap is opened even for a very small value of $U$, as the similar case for the Hubbard model without the superlattice potential \cite{PhysRevLett.20.1445}. The charge gap rises rapidly as $U$ increases, and eventually tends to a finite value. For the Hermitian case, the charge gap tends to $\Delta_b/2$ for large $U$ \cite{PhysRevB.88.045110}, where $\Delta_b$ is the single-particle gap  at the $1/3$ particle filling; while the saturated charge gap in the non-Hermitian case is well below this value ($\Delta_b/2$ = 0.54) due to the nonreciprocal hopping.

\subsection{Many-body spectrum}
To address whether the non-Hermitian Mott insulator is topologically nontrivial, we calculate the low-energy spectra of neutral excitations in the many-excitation subspace considering both periodic and open boundaries, as summarized in Fig.~\ref{fig2}. For the weak repulsive interaction, e.g., $U = 1$, there are no gaps for both periodic [see Fig.~\ref{fig2}(a)] and open [see Fig.~\ref{fig2}(b)] boundaries, and no gapless neutral excitations are thus observed. However, in the strong repulsive limit, e.g., $U = 10$ and $U = 100$, the lower-energy sector and higher-excited levels cross at $\phi = 2\pi/3$ in the bulk gap regime, and gapless neutral excitations emerge for open boundaries [see Figs.~\ref{fig2}(d,f)]. Note that the lower-energy sector contains six energy levels (i.e., six kinds of spin configurations) and will become degenerate in the infinite-$U$ limit. Its degeneracy is lifted for a finite value due to spin fluctuations. For periodic-system energy spectra in Figs.~\ref{fig2}(c,e), even in the large-$U$ limit, no gap is opened between the lower-energy sector and higher-excited levels, due to the strong nonreciprocal hopping with $\alpha=0.4$ (see also the eigenenergies on the complex plane in Fig.~\ref{FigN1}(a)]).  This indicates that the energy spectrum of the non-Hermitian interacting system is sensitive to the boundary conditions, which is a manifestation of the \textit{non-Hermitian skin effect}. 

In contrast to the case of the strong nonreciprocal hopping, the periodic-system energy spectrum is gapped for the weak nonreciprocal hopping  with $\alpha=0.2$, as shown in Fig.~\ref{FigN1}(b) and Fig.~\ref{figNS2}. This results from the interplay of non-Hermiticity and superlattice potential (see also the single-particle spectra in Appendix \ref{Appendix_B}). In addition, the energy spectrum is always real for open boundaries. This can be seen via the similarity transformations 
\begin{align}\label{p1_H1121}
c_{j,\sigma} \rightarrow e^{j\alpha} c_{j,\sigma},~~~ c_{j,\sigma}^\dagger \rightarrow e^{-j\alpha} c_{j,\sigma}^\dagger, 
\end{align}
which map the non-Hermitian Hamiltonian $\mathcal{H}$ in Eq.~(\ref{p1_H11}) onto the Hermitian one. 

To explore the edge localization of the gapless neutral excitation, we calculate its spatial charge and spin distributions. Note that the non-Hermitian Hamiltonian holds different left and right eigenvectors, which are defined as $\mathcal{H}^\dagger \ket{\Phi}_L = E^* \ket{\Phi}_L$, and $\mathcal{H} \ket{\Phi}_R = E \ket{\Phi}_R$. They satisfy the biorthogonal normalization condition ${}_L\bra{\Phi} \ket{\Phi}_R = 1$. As such, the spatial charge and spin distributions can be calculated via either biorthogonal eigenvectors or only right eigenvectors. In order to detect the non-Hermitian skin effect, as observed in the single-particle model, we adopt the right eigenvectors, where the spatial charge and spin distributions are defined as 
\begin{align}\label{p1_H7}
	\Delta n_{RR,j}^{\textrm{ne}} = & \frac{{}_R\bra{\Phi_1 \left(N_\uparrow, N_\downarrow\right)} n_j \ket{\Phi_1 \left(N_\uparrow, N_\downarrow\right)}_R}{{}_R\bra{\Phi_1 \left(N_\uparrow, N_\downarrow\right)}\ket{\Phi_1 \left(N_\uparrow, N_\downarrow\right)}_R} -\nonumber \\
	&  \frac{{}_R\bra{\Phi_0 \left(N_\uparrow, N_\downarrow\right)} n_j \ket{\Phi_0 \left(N_\uparrow, N_\downarrow\right)}_R}{{}_R\bra{\Phi_0 \left(N_\uparrow, N_\downarrow\right)}\ket{\Phi_0 \left(N_\uparrow, N_\downarrow\right)}_R}, 
\end{align}
\begin{align}\label{p1_H8}
	\Delta S_{RR,j}^{\textrm{ne}} = & \frac{{}_R\bra{\Phi_1 \left(N_\uparrow, N_\downarrow\right)} S_j \ket{\Phi_1 \left(N_\uparrow, N_\downarrow\right)}_R}{{}_R\bra{\Phi_1 \left(N_\uparrow, N_\downarrow\right)}\ket{\Phi_1 \left(N_\uparrow, N_\downarrow\right)}_R} -\nonumber \\
	&  \frac{{}_R\bra{\Phi_0 \left(N_\uparrow, N_\downarrow\right)} S_j \ket{\Phi_0 \left(N_\uparrow, N_\downarrow\right)}_R}{{}_R\bra{\Phi_0 \left(N_\uparrow, N_\downarrow\right)}\ket{\Phi_0 \left(N_\uparrow, N_\downarrow\right)}_R},
\end{align}
where $n_j = n_{j,\uparrow} + n_{j,\downarrow} $, $S_{j} = \left(n_{j,\uparrow} - n_{j,\downarrow}\right)/2$,  the superscript ``ne'' refers to neutral, and $\ket{\Phi_0 \left(N_\uparrow, N_\downarrow\right)}_R$ ($\ket{\Phi_1 \left(N_\uparrow, N_\downarrow\right)}_R$) is the right lowest (higher) energy state. 

Equations (\ref{p1_H7}) and (\ref{p1_H8}) provide the differences of density and magnetization distributions of two in-gap modes in the upper and lower branches. We plot the edge excitations and ground-state population distributions of the neutral excitations for $U=100$ in Fig.~\ref{fig3}. According to Figs.~\ref{fig3}(a,b), the gapless neutral excitation only carries a charge degree of freedom at the edges. Moreover, only the in-gap states with $\phi < 2\pi/3$ exhibit edge excitations for $\alpha = 0.4$ and $V_0=1.5$, as shown in Fig.~\ref{fig3}(a). In contrast, for $\alpha = -0.4$, edge excitations emerge only if $\phi > 2\pi/3$, as shown in Fig.~\ref{fig3}(c). Note that the absence of edge excitations also appears for the weak interaction limit, e.g., for $U=10$ in Appendix \ref{Appendix_C}. Therefore, the charge distributions calculated by only the right eigenvectors indicate that the non-Hermitian interacting system shows the anomalous boundary effect, where some in-gap states exhibit no edge excitations in the open chain. However, this phenomenon is absent for the non-Hermitian single-particle case (see Appendix \ref{Appendix_B}). 

Moreover, such an anomalous boundary effect disappears for weak nonreciprocal hopping, i.e., $\alpha=0.2$, as shown in Fig.~\ref{fig3}(d), or large modulation amplitude, i.e., $V_0=4$ in Fig.~\ref{fig3}(e). These indicate that \textit{the anomalous boundary effect results from the combined effects of the nonreciprocal hopping, superlattice potential, and interactions.} The interactions drive the metallic phase into a topological insulator,  the superlattice potential forces particles to occupy the site with lower potential in the ground state, and the nonreciprocal hopping pushes particles accumulated towards one of two ends in the higher-excited state. These lead to the absence of neutral excitations at the edges for certain values of $\phi$. Furthermore, unlike the single-particle model (see Appendix \ref{Appendix_B}), the nonreciprocal hopping here does not cause bulk charges to accumulate near the boundaries due to the Pauli exclusion principle and interactions [see Fig.~\ref{fig3}(f)].  

To explain the anomalous boundary effect, we plot the spin-resolved charge densities calculated by only the right eigenvectors, as shown in Fig.~\ref{figNS4}. The charge densities in the first row are calculated with the higher-excited state $\ket{\Phi_1 \left(N_\uparrow, N_\downarrow\right)}_R$, and  the second row with the ground state $\ket{\Phi_0 \left(N_\uparrow, N_\downarrow\right)}_R$  for $\alpha=0.4$ and $\alpha=0$. Their difference defines the corresponding charge distributions of the neutral excitations [see Eq.~(\ref{p1_H7})]. The black dotted line indicates the superlattice potential $V_j = V_0 \cos(2 \pi j/q  + \phi)$ for $\phi=2\pi/5$ and $\phi=4\pi/5$. 

For the Hermitian case with $\alpha=0$, due to the spatial modulations of the superlattice potential, the charge densities in the bulk exhibit periodic oscillations, and the neutral excitations are localized at the ends, for both $\phi=2\pi/5$ and $\phi=4\pi/5$, as indicated in Figs.~\ref{figNS4}(c,d).  

For the non-Hermitian case with $\alpha=0.4$, the charge densities are also periodically modulated by the superlattice potential. The particles in the bulk do not accumulate towards the right end, in spite of the stronger forward-hopping amplitude, due to the Pauli exclusion principle and on-site interactions. In order to minimize the energy in the ground state $\ket{\Phi_0 \left(N_\uparrow, N_\downarrow\right)}_R$ (the second row), the site where the superlattice potential $V_j$ is lower is occupied by more charges. As shown in Figs.~\ref{figNS4}(a,c) and Figs.~\ref{figNS4}(b,d), to minimize the ground-state energy, the left end is occupied by more particles than the right end for $\phi=2\pi/5$; while the right end is occupied by more particles than the left end for $\phi=4\pi/5$. In the excited state $\ket{\Phi_1 \left(N_\uparrow, N_\downarrow\right)}_R$ (the first row), the nonreciprocal hopping somehow pushes the particles towards the right end both for $\phi=2\pi/5$ and $\phi=4\pi/5$, as shown in Figs.~\ref{figNS4}(a,b). As a result, the neutral excitation is still localized at the ends for $\phi=2\pi/5$ and $\alpha=0.4$, but the edge neutral excitations disappear for $\phi=4\pi/5$ and $\alpha=0.4$. 

Moreover, when the modulation potential increases from the $V_0=1.5$ to $V_0=4$ [see Fig.~\ref{fig3}(e)], the nonreciprocal hopping is weakened by the potential barrier for the higher-energy states; therefore, the neutral excitation is localized at ends for $\phi=4\pi/5$ and $V_0=4$. Therefore, the anomalous boundary effect of the neutral excitation results from the combined effects of nonreciprocal hopping, superlattice modulation, and interactions. 

The charge distributions computed by only the right eigenvectors lead to the absences of edge excitations for some in-gap states due to the intrinsic non-Hermitian skin effect, as discussed above. To correctly characterize the edge excitations, we resort to the biorthogonal formula. The charge distribution in biorthogonal eigenvectors is calculated as 
\begin{align}\label{p1_H9}
	\Delta n_{LR,j}^{\textrm{ne}} = & {}_L\bra{\Phi_1 \left(N_\uparrow, N_\downarrow\right)} n_j \ket{\Phi_1 \left(N_\uparrow, N_\downarrow\right)}_R -\nonumber \\
	&  {}_L\bra{\Phi_0 \left(N_\uparrow, N_\downarrow\right)} n_j \ket{\Phi_0 \left(N_\uparrow, N_\downarrow\right)}_R. 
\end{align}
As shown in Fig.~\ref{fig32}, all the in-gap states of the neutral excitations are localized at the edges. In addition, the charge distributions with $\alpha = 0.4$ [see Fig.~\ref{fig32}(a)] are the same as the ones with $\alpha = -0.4$ [see Fig.~\ref{fig32}(b)], indicating that they are insensitive to the nonreciprocal hopping based on the biorthogonal formula.

\subsection{Quasiparticle spectrum}
To further explore the topological properties of the Mott insulator, we proceed to calculate the quasiparticle energy spectrum, or charge excitation spectrum, defined as 
\begin{align}\label{p1_H110}
\Delta E_c(N_\uparrow, N_\downarrow) = E_0\left(N_\uparrow, N_\downarrow+1 \right) - E_0\left(N_\uparrow, N_\downarrow \right).
\end{align}
The corresponding spatial charge distribution, calculated by both only the right eigenvectors and biorthogonal eigenvectors, are: 
\begin{align}\label{p1_H10}
	\Delta n_{RR,j}^{c} = & \frac{{}_R\bra{\Phi_0 \left(N_\uparrow, N_\downarrow+1\right)} n_j \ket{\Phi_0 \left(N_\uparrow, N_\downarrow+1\right)}_R}{{}_R\bra{\Phi_0 \left(N_\uparrow, N_\downarrow+1\right)}\ket{\Phi_0 \left(N_\uparrow, N_\downarrow+1\right)}_R} -\nonumber \\
	&  \frac{{}_R\bra{\Phi_0 \left(N_\uparrow, N_\downarrow\right)} n_j \ket{\Phi_0 \left(N_\uparrow, N_\downarrow\right)}_R}{{}_R\bra{\Phi_0 \left(N_\uparrow, N_\downarrow\right)}\ket{\Phi_0 \left(N_\uparrow, N_\downarrow\right)}_R}, 
\end{align}
\begin{align}\label{p1_H12}
	\Delta n_{LR,j}^{c} = & {}_L\bra{\Phi_0 \left(N_\uparrow, N_\downarrow+1\right)} n_j \ket{\Phi_0 \left(N_\uparrow, N_\downarrow+1\right)}_R -\nonumber \\
	&  {}_L\bra{\Phi_0 \left(N_\uparrow, N_\downarrow\right)} n_j \ket{\Phi_0 \left(N_\uparrow, N_\downarrow\right)}_R, 
\end{align}

Figures \ref{fig4}(a,b) plot the charge excitation spectra versus the modulation phase $\phi$ under both periodic and open boundaries for $\Delta E_c(N_\uparrow=2, N_\downarrow=2)$ and $\Delta E_c(N_\uparrow=2, N_\downarrow=1)$. For periodic boundary, the charge excitation spectrum is gapped. Once the boundary is opened, the in-gap modes appear. 

To explore the non-Hermitian skin effect on edge charge excitations, Figure ~\ref{fig4}(c,d,e) shows the spatial charge distributions calculated using only the right eigenvector. As the same case of neutral excitations, only the in-gap modes of the charge excitations with $\phi < 2\pi/3$ are localized at the edges for the strong nonreciprocal hopping, i.e., $\alpha = 0.4$ in Fig.~\ref{fig4}(c), while the edge charge excitations disappears for $\phi > 2\pi/3$. When the amplitude of the nonreciprocal hopping $\alpha$   is reduced [see Fig.~\ref{fig4}(d)], or the modulation intensity $V_0$  is increased [see Fig.~\ref{fig4}(e)], the edge excitations of the in-gap modes for both $\phi < 2\pi/3$ and $\phi > 2\pi/3$ are observed. \textit{This anomalous boundary effect for the edge charge excitations is due to the combined results of nonreciprocal hopping, superlattice potential, and interactions.} 

To explain such an anomalous boundary effect for the edge charge excitations, we plot the charge densities of the ground state $\ket{\Phi_0 \left(N_\uparrow, N_\downarrow\right)}_R$ (the  first row) and $\ket{\Phi_0 \left(N_\uparrow, N_\downarrow+1\right)}_R$ (the second row), as shown in Fig.~\ref{figNS6}.  Their difference defines the corresponding charge distributions of the charge excitations [see Eq.~(\ref{p1_H10})]. Due to the spatial modulations of the superlattice potential, the charge densities in the bulk exhibit periodic oscillations. To minimize the system energy in the ground state, the site where the superlattice potential $V_j$ is lower is occupied by more charges. Therefore, in the ground state $\ket{\Phi_0 \left(N_\uparrow, N_\downarrow\right)}_R$ (the  first row in Fig.~\ref{figNS6}), the left end is occupied by more particles than the right end for $\phi=2\pi/5$, while the right end is occupied by more particles than the left end for $\phi=4\pi/5$. When an extra particle is added, in the ground state $\ket{\Phi_0 \left(N_\uparrow, N_\downarrow+1\right)}_R$ (the second row in Fig.~\ref{figNS6}), it tends to occupy the right end for $\alpha=0.4$ due to the stronger forward hopping. This leads to the absence of the charge excitations for $\phi=4\pi/5$ and $\alpha=0.4$. 

To fully characterize boundary charge excitations, we can implement biorthogonal eigenvectors, where all the in-gap modes show edge excitations, as shown in Fig.~\ref{fig4}(f).

\begin{figure}[!b]
	\centering
	\includegraphics[width=7.0cm]{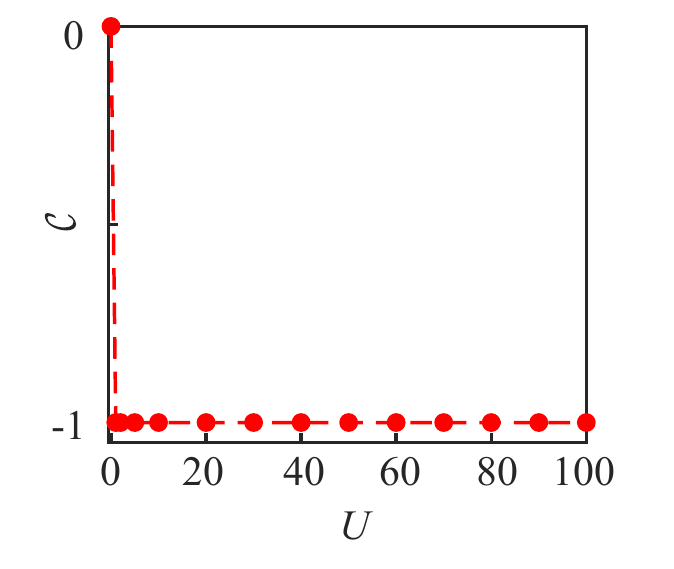}
	\caption{Chern number $\mathcal{C}$ versus $U$, with parameters $L = 18$, $N_\uparrow = 3$, $N_\downarrow = 3$, $t=1$, $V_0=1.5$, $q = 3$, $\phi = 2 \pi/3$, and $\alpha = 0$.}\label{fig15}
\end{figure}
%

%
\begin{figure*}[!tb]
	\centering
	\includegraphics[width=18cm]{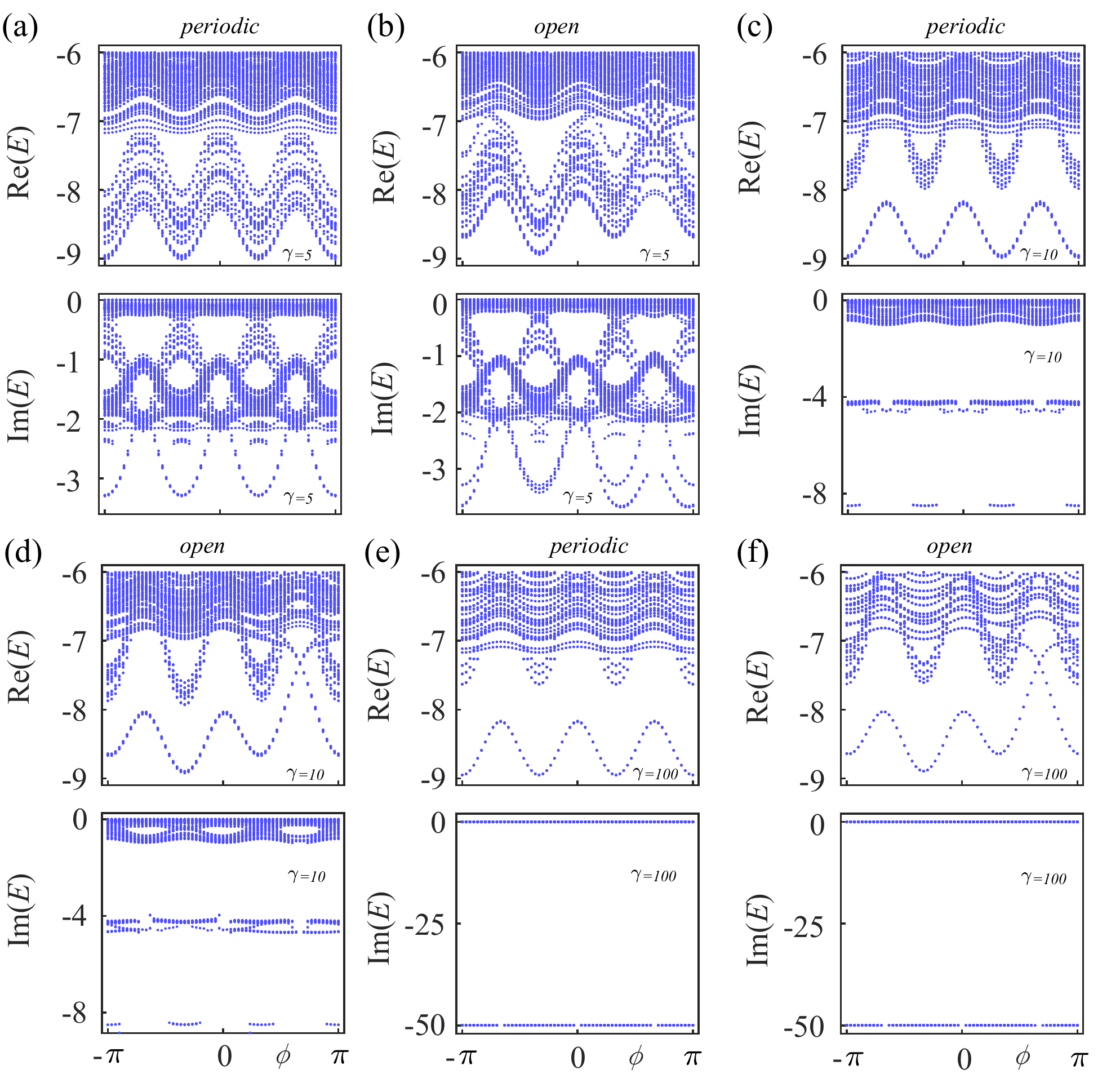}
	\caption{Low-energy spectrum versus the modulation phase $\phi$ for: (a,b)  $\gamma = 5$; (c,d) $\gamma = 10$; and (e,f) $\gamma = 100$. The spectra in (a,c,e) are for periodic boundary, and in (b,d,f) using open boundaries. The parameters used here are: $L = 12$, $N_\uparrow = 2$, $N_\downarrow = 2$, $t=1$, $V_0=1.5$, $q=3$, and $U = 0$.}\label{FigSM1}
\end{figure*}
%

\subsection{Topological invariant}
To characterize the topological feature of the system considered, we compute the many-body Chern number, which is defined as: 
\begin{align}\label{p3_H1}
	\mathcal{C} = \frac{1}{2 \pi} \int_{-\pi}^{\pi} d\theta \int_{-\pi}^{\pi} d\phi \left(\partial_\theta A_\phi - \partial_\phi A_\theta\right), 
\end{align}
where the Berry connection $A_\mu = i {}_L\bra{\Psi_g} \partial_\mu \ket{\Psi_g}_R$ ($\mu = \theta,~\phi$). $\ket{\Psi_g}_R$ ($\ket{\Psi_g}_L$) is the many-body right (left) ground state under the twist boundary conditions $c_{j+L,\sigma} = e^{i\theta} c_{j,\sigma}$, with twist angle $\theta$. Since the low-energy spectrum of the neutral excitation is gapless even for a strong repulsive limit due to the nonreciprocal hopping, the Chern number based on the Hamiltonian $\mathcal{H}$ in Eq.~(\ref{p1_H11}) fails to characterize the bulk-boundary correspondence due to the sensitivity of many-body spectra to boundary conditions. We note that the non-Hermitian Hamiltonian $\mathcal{H}$ can be mapped onto a Hermitian one via a similarity transformation, and they share the same quasiparticle bands for open boundaries. Thus, we can restore the bulk-boundary correspondence by computing the Chern number of the corresponding Hermitian Hamiltonian under its twisted boundary condition. Figure \ref{fig15} plots the topological phase diagram with $\alpha=0$, where $\mathcal{C} = -1$ even for very small $U$, because an arbitrary small repulsive interaction [see also Fig.~\ref{fig1}(b)] can drive the Hubbard system into topological Mott phases \cite{PhysRevB.88.045110,PhysRevLett.20.1445}.

\section{Non-Hermitian model with complex-valued interaction}

\subsection{Effective Hamiltonian}
We now discuss the strongly correlated topological phases with complex-valued interactions. We consider 1D spin-1/2 ultracold fermionic atoms loaded in a bichromatic optical lattice, described by $\mathcal{H} = \mathcal{H}_0 + \mathcal{H}_\textrm{int}$ with
\begin{align}\label{p1_H1}
	\mathcal{H}_0 = -t \sum_{j,\sigma} \left(c_{j,\sigma}^\dagger c_{j+1,\sigma} + c_{j+1,\sigma}^\dagger c_{j,\sigma}\right) + \sum_{j,\sigma} V_j n_{j,\sigma}, 
\end{align}
\begin{align}\label{p1_H2}
	\mathcal{H}_\textrm{int} = U \sum_{j} c_{j,\uparrow}^\dagger c_{j,\downarrow}^\dagger c_{j,\downarrow} c_{j,\uparrow}, 
\end{align}
where $V_j = V_0 \cos(2 \pi \beta j  + \phi)$.

Now we consider two-body inelastic collisions between ultracold atoms, which leads to losses of atoms \cite{Tomita2017}. The dissipative dynamics of the system is described by the following quantum master equation \cite{Breuer2007}  
\begin{align}\label{p1_H3}
	\frac{d \rho}{dt} = & -i \left[\mathcal{H}, ~\rho\right] -\frac{\gamma}{2} \sum_j \left(L_j^\dagger L_j \rho + \rho L_j^\dagger L_j - 2 L_j \rho L_j^\dagger\right) \nonumber \\
	=& -i \left[\mathcal{H}_{\textrm{eff}}, ~\rho\right] + \gamma \sum_j  L_j \rho L_j^\dagger,
\end{align}
where $\rho$ is the density matrix of the ultracold atoms, and the Lindblad operator $L_j = c_{j,\downarrow} c_{j,\uparrow}$ describes a two-particle loss at site $j$ with rate $\gamma$. According to the quantum-trajectory theory \cite{Meystre2007,PhysRevLett.68.580,PhysRevLett.70.2273}, the dynamics of the dissipative quantum system described by the first line in Eq.~(\ref{p1_H3}) can be decomposed into two processes: a non-unitary Schr$\ddot{\textrm{o}}$dinger evolution described by the effective non-Hermitian Hamiltonian $\mathcal{H}_{\textrm{eff}} = \mathcal{H} - i \gamma/2 \sum_{j} L_j^\dagger L_j$, and stochastic quantum jumps, described by the second term in the second line of Eq.~(\ref{p1_H3}), which leads to atomic losses. Therefore, when projecting out the quantum jumps (by continuously monitoring the particle number \cite{PhysRevA.94.053615, Ashida2017}), the system is governed by the following effective non-Hermitian Hamiltonian with a complex-valued on-site interaction  
\begin{align}\label{p1_H4}
	\mathcal{H}_{\textrm{eff}} = \mathcal{H}_0 + \left(U - i \gamma/2\right) \sum_{j} c_{j,\uparrow}^\dagger c_{j,\downarrow}^\dagger c_{j,\downarrow} c_{j,\uparrow}.
\end{align}

\begin{figure}[!tb]
	\centering
	\includegraphics[width=8.4cm]{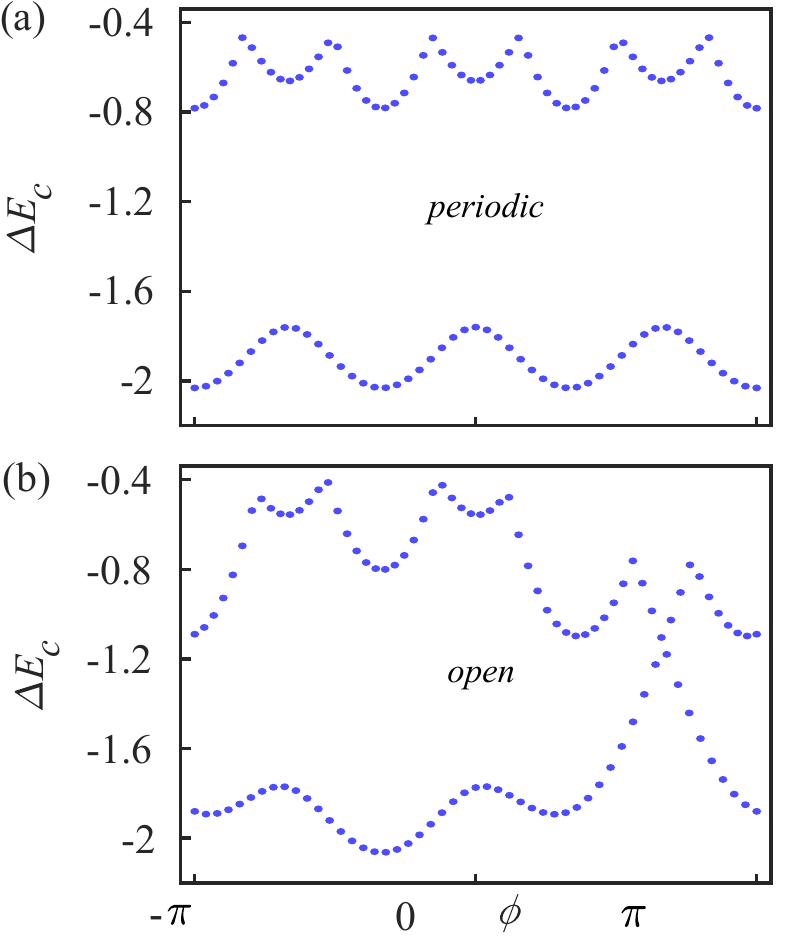}
	\caption{Charge excitation spectra $\Delta E_c(2, 2)$ and $\Delta E_c(2, 1)$ versus the modulation phase $\phi$ for  (a) periodic and (b) open boundaries. The parameters used here are:  $\gamma = 100$, $L = 12$, $N_\uparrow = 2$, $N_\downarrow = 2$, $t=1$, $V_0=1.5$, $q=3$, and $U = 0$.}\label{FigSM2}
\end{figure}

\begin{figure}[!tb]
	\centering
	\includegraphics[width=8.4cm]{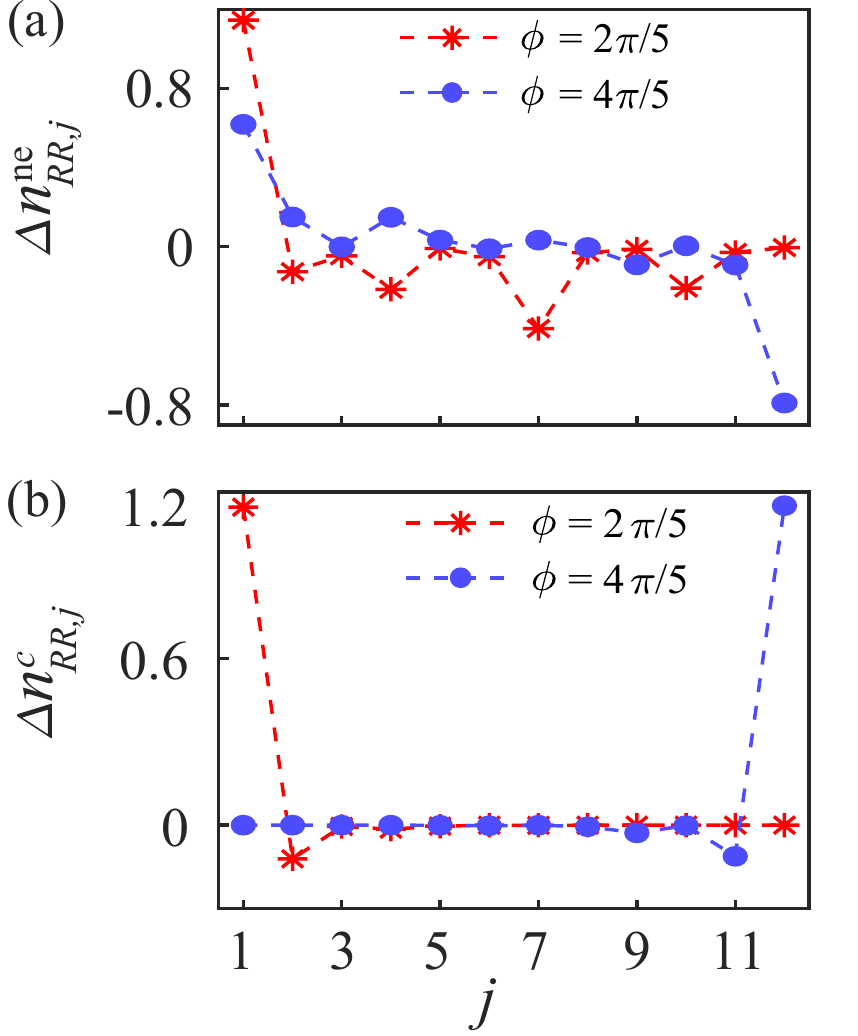}
	\caption{Spatial charge distributions for the (a) neutral excitation and (b) charge excitation modes for different modulation phases $\phi$. The parameters used here are:  $\gamma = 100$, $L = 12$, $N_\uparrow = 2$, $N_\downarrow = 2$, $t=1$, $V_0=1.5$, $q=3$, and $U = 0$.}\label{fig11}
\end{figure}

\subsection{Low-energy spectra of neutral and charge excitations}
To address whether the complexed-value interaction can drive the metallic phases into topologically nontrivial insulators, we calculate the low-energy spectra of neutral excitations for the purely imaginary-valued interactions using both periodic and open boundary conditions, as shown in Fig.~\ref{FigSM1}. For a small absolute value of the interaction strength, e.g., $\gamma=5$, there exist no gaps using both periodic [Fig.~\ref{FigSM1}(a)] and open [Fig.~\ref{FigSM1}(b)] boundary conditions.  However, for large absolute values of the interaction strengths, e.g., $\gamma=10$ and $\gamma=100$, the energy gaps are opened for periodic boundary condition, as shown in Figs.~\ref{FigSM1}(c) and \ref{FigSM1}(e). The gap opening indicates that the purely imaginary-valued on-site interaction here plays the roles of the effective repulsion between the two-component fermions, which can be attributed to a phenomenon
similar to the continuous quantum Zeno effect \cite{Syassen2008, PhysRevLett.110.035302, Tomita2017, PhysRevLett.123.123601}. The strong two-body inelastic collision, acting as the strong measurement,  effectively suppresses the coherent tunneling, thereby preventing double occupancies of fermions with opposite spin components. When the system goes from periodic to open boundaries, the in-gap modes, which connect the lower-energy  and higher-excited sectors, appear, as shown in Figs.~\ref{FigSM1}(d) and \ref{FigSM1}(f). The gapless modes here closely resemble the appearance of edge states in the single-particle spectrum. In contrast to the nonreciprocal case, the many-body spectrum is complex. Note that the lower-energy sector become broader for smaller interaction strengths due to spin fluctuations.

In addition to the neutral excitations, we present the low-energy spectra of charge excitations for the purely imaginary-valued interactions using both periodic and open boundary conditions, as shown in Fig.~\ref{FigSM2}. When the 1D chain is under the periodic boundary, the spectrum of charge excitation is gaped for large absolute values of the interaction strengths, e.g., $\gamma=100$ in Fig.~\ref{FigSM2}(a). Once its boundary is opened, the gapless edge excitations emerge, as shown in Fig.~\ref{FigSM2}(b). 

Both neutral and charge excitation spectra support in-gap modes. Their spatial charge distributions calculated by right eigenvectors are plotted in Figs.~\ref{fig11}(a,b). For both cases, the in-gap modes are well localized at the edges. Note that these exhibit the same spatial distributions calculated by biorthogonal eigenvectors, for the neutral and charge excitations, as the one using only right eigenvectors. No anomalous boundary effect occurs for purely imaginary-valued interactions.

\section{Summary and discussion}
We have discussed topological properties of an interaction-induced topological Mott insulator in a 1D non-Hermitian spinful fermionic superlattice system. We analyzed its low-energy neutral and charge excitations spectra in the presence of nonreciprocal hopping, where in-gap modes appear. We found that the nonreciprocal hopping makes the neutral excitation spectrum sensitive to the boundary conditions, which is a manifestation of the non-Hermitian skin effect. However, the unique non-Hermitian skin effect on particle density, as shown at the single-particle case, is absent due to the Pauli exclusion principle. The interplay of nonreciprocal hopping, superlattice potential, and interactions leads to the absence of edge excitations, defined by only the right eigenvectors, of some in-gap modes appearing in both the neutral and charge excitation spectra. Moreover, these edge excitations can be characterized by using biorthogonal eigenvectors. Furthermore, the topological Mott insulator induced by purely imaginary-valued interaction can be interpreted by the continuous quantum Zeno effect.  

The non-Hermitian skin effect has been shown to cause unusual properties in fermionic many-body systems, i.e., a Fermi surface in real space \cite{arXiv:1911.00023}. Our findings indicate that the non-Hermitian skin effect, in combination with a periodic modulation and interactions, can lead to unconventional topological features without Hermitian counterparts, which is worth further  exploration. Possible future research directions include non-Hermitian fractional quantum Hall effect \cite{PhysRevLett.110.215301}, fractional charge pumping \cite{PhysRevB.94.235139}, and an extension to higher-dimensional systems \cite{Lohse2018}. Meanwhile, we hope that our studies inspire further future explorations of the role of the non-Hermitian skin effect and searching for novel topological features in non-Hermitian interacting fermionic systems. 

\textit{Noted added}: After this work was completed, we became aware of two related works \cite{arXiv:2001.07088, arXiv:2002.00554}, which discuss non-Hermitian topological Mott insulators of interacting bosons. Unlike the fermionic model considered here, the nonreciprocal bosonic model exhibits the accumulation of particle density at the edges, and no anomalous boundary effect is reported \cite{arXiv:2001.07088}.

\begin{acknowledgments}
T.L. is grateful to Zhong Wang, Haiping Hu, Clemens Gneiting and Zhihao Xu for valuable discussions. T.L. acknowledges support from the Grant-in-Aid for a JSPS Foreign Postdoctoral Fellowship (P18023). T.Y. is supported by JSPS KAKENHI (Grants No.~JP19K21032). Z.L.X thanks the support from the NSFC (Grant No.~11874432). F.N. is supported in part by: NTT Research,
Army Research Office (ARO) (Grant No.~W911NF-18-1-0358),
Japan Science and Technology Agency (JST)
(via the CREST Grant No.~JPMJCR1676),
Japan Society for the Promotion of Science (JSPS) (via the KAKENHI Grant No.~JP20H00134
and the JSPS-RFBR Grant No.~JPJSBP120194828),
the Asian Office of Aerospace Research and Development (AOARD),
and the Foundational Questions Institute Fund (FQXi) via Grant No.~FQXi-IAF19-06.
\end{acknowledgments}

%
\begin{figure*}[!tb]
	\centering
	\includegraphics[width=18cm]{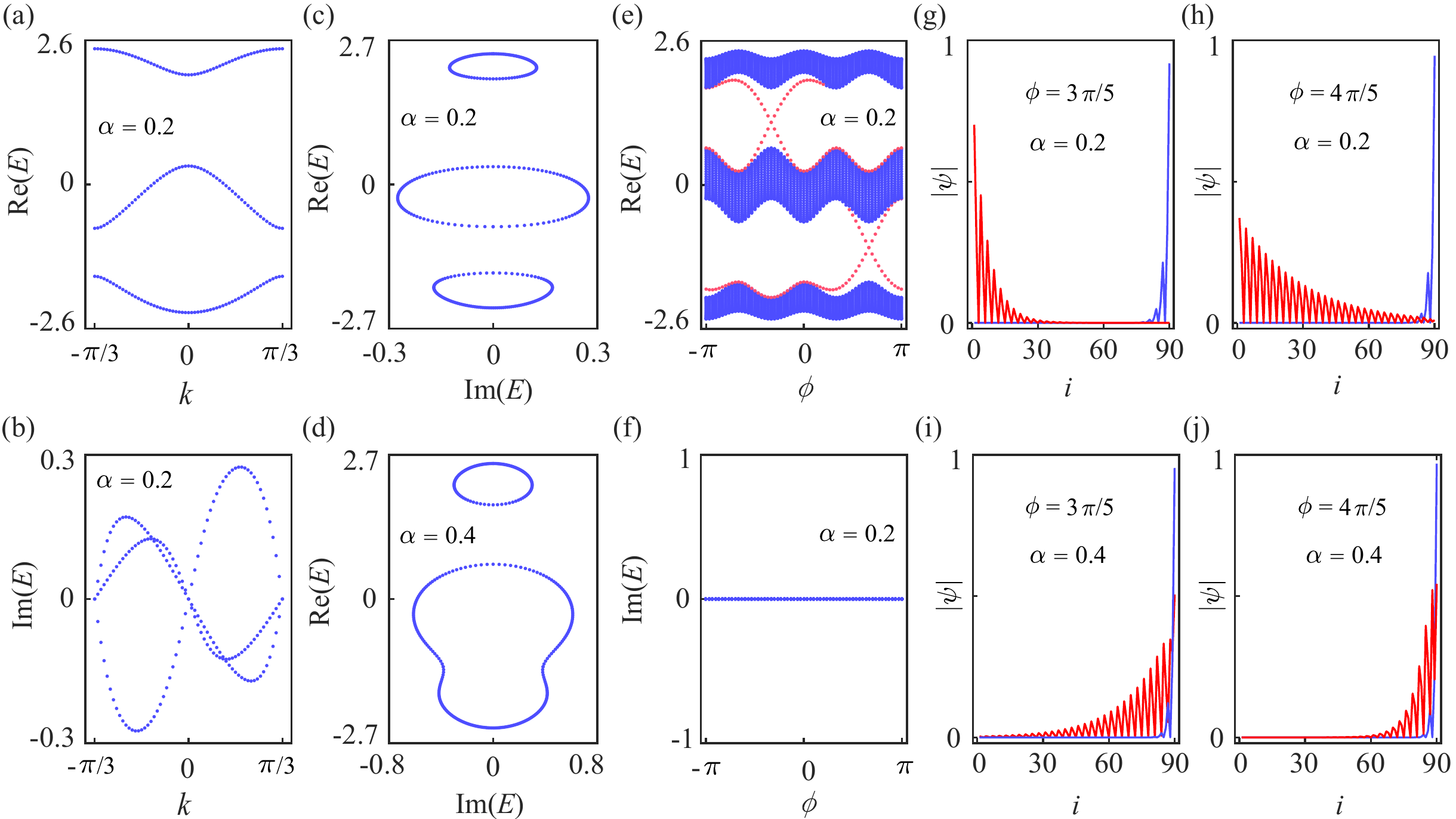}
	\caption{Non-Hermitian single-particle spectra. (a) Real part and (b) imaginary parts of complex eigenenergies for $\alpha = 0.2$ using the periodic boundary. Real part vs. imaginary part of complex eigenenergies (c) for $\alpha = 0.2$ and (d) for $\alpha = 0.4$ considering the periodic boundary. (e) Real part and (f) imaginary parts of complex eigenenergies as a function of the modulation phase $\phi$ for $\alpha = 0.2$ using the open boundaries. Red doted curves indicate the edge modes. (g-j) Spatial density distributions of two mid-gap states at a specific modulation phase $\phi$ and nonreciprocal hopping factor $\alpha$. The parameters used here are: $L = 90$, $t=1$, $V_0=1.5$, and $q=3$.}\label{FigNS1}
\end{figure*}
%

%
\begin{figure*}[!tb]
	\centering
	\includegraphics[width=18cm]{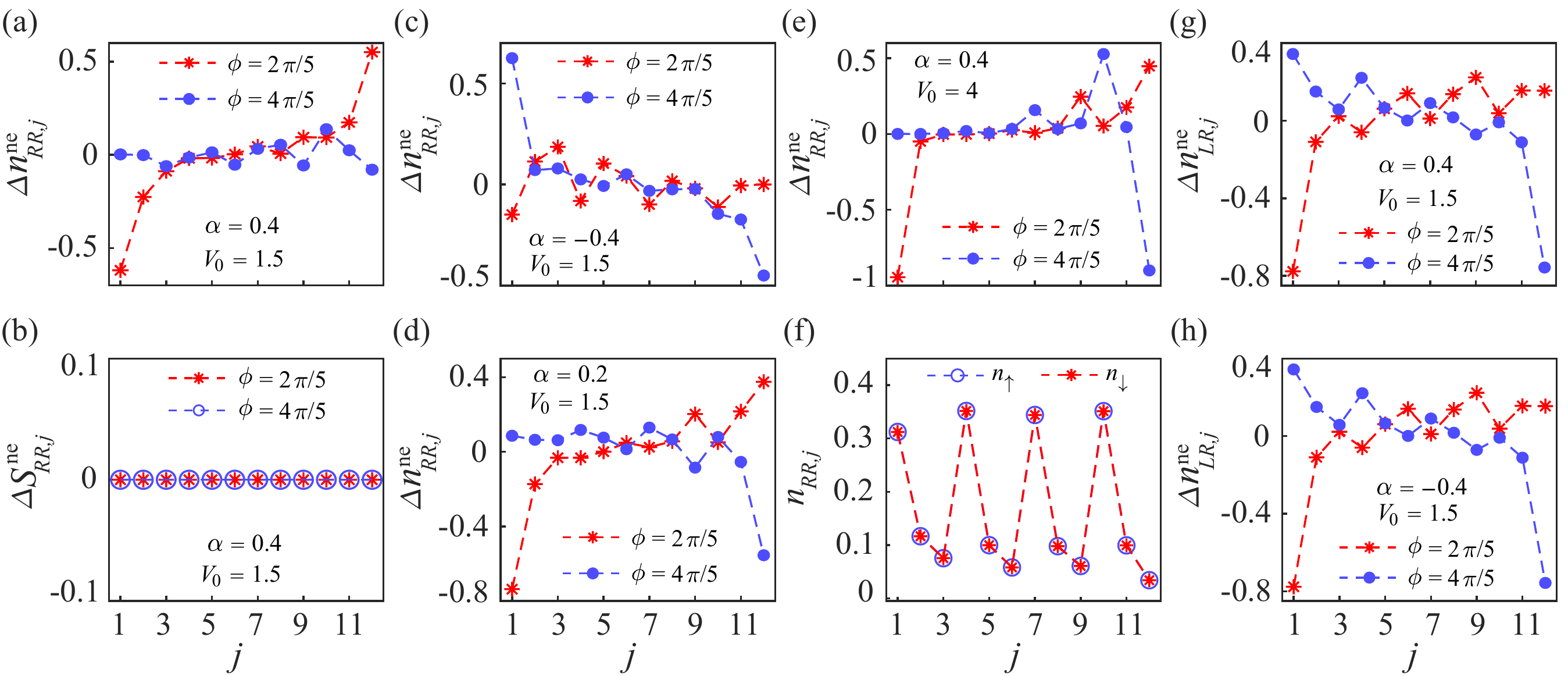}
	\caption{Spatial distributions of the (a,c,d,e) charge and (b) spin for the neutral excitation modes, calculated via only the right eigenvectors, for different modulation phases $\phi$. The spatial distributions are calculated for: (a,b) $\alpha = 0.4$, $V_0 = 1.5$; (c) $\alpha = -0.4$, $V_0 = 1.5$; (d)  $\alpha = 0.2$, $V_0 = 1.5$; and (e) $\alpha = 0.4$, $V_0 = 4$. (f) Spin-resolved charge densities of the ground state for $\phi = 2\pi/5$, $\alpha = 0.4$ and $V_0 = 1.5$. The spatial charge distributions calculated via biorthogonal eigenvectors for: (g) $\alpha=0.4$, $V_0 = 1.5$; and (h) $\alpha=-0.4$;$V_0 = 1.5$. The parameters used here are: $L = 12$, $N_\uparrow = 2$, $N_\downarrow = 2$, $t=1$, $q = 3$, $V_0=1.5$, and $U = 10$.}\label{figNS3}
\end{figure*}

\appendix
\section{Experimental realization of Fermi-Hubbard model with nonreciprocal hopping}
\label{Appendix_A}

To experimentally realize the non-Hermitian Fermi-Hubbard model with  asymmetric hopping in the ultracold atom systems, we can employ  reservoir engineering \cite{PhysRevX.8.031079}. We rewrite the system Hamiltonian Eq.~(\ref{p1_H11}) into Hermitian $\mathcal{H}_1 = \left(\mathcal{H} + \mathcal{H}^\dagger\right)/2$, and anti-Hermitian $\mathcal{H}_2 = \left(\mathcal{H} - \mathcal{H}^\dagger\right)/2$ parts:
\begin{align}\label{p1}
	\mathcal{H}_1 = &-\frac{t\left(e^{\alpha}+e^{-\alpha}\right)}{2} \sum_{j,\sigma} \left(  c_{j+1,\sigma}^\dagger c_{j,\sigma} + c_{j,\sigma}^\dagger c_{j+1,\sigma}\right) \nonumber \\
	&+ \sum_{j,\sigma} V_j n_{j,\sigma}  + U \sum_{j} c_{j,\uparrow}^\dagger c_{j,\downarrow}^\dagger c_{j,\downarrow} c_{j,\uparrow},  
\end{align}
\begin{align}\label{p2}
	\mathcal{H}_2 = -\frac{t\left(e^{\alpha}-e^{-\alpha}\right)}{2} \sum_{j,\sigma} \left( c_{j+1,\sigma}^\dagger c_{j,\sigma} - c_{j,\sigma}^\dagger c_{j+1,\sigma}\right),  
\end{align}
To engineer this anti-Hermitian part $\mathcal{H}_2$, we follow the method developed in Ref.~\cite{PhysRevX.8.031079} to dissipatively  engineer the following jump operators that describe the collective loss of two nearest-neighbor sites:
\begin{align}\label{p3}
	L_j = \sqrt{t\left(e^{\alpha}-e^{-\alpha}\right)\textrm{sgn}(\alpha)} \left[c_{j,\sigma} + i\textrm{sgn}(\alpha) c_{j+1,\sigma}  \right],  
\end{align}
where $\textrm{sgn}$ refers to $\pm$ signs. Then, the open quantum system dynamics with post-selection measurements is determined by the non-Hermitian effective Hamiltonian:
\begin{align}\label{p4}
	H_{\textrm{eff}} & =  \mathcal{H} - \frac{i}{2}\sum_j L_j^\dagger L_j 
	\nonumber \\& = \mathcal{H} - i\frac{t\left(e^{\alpha}-e^{-\alpha}\right)\textrm{sgn}(\alpha)}{2} \sum_j c_{j,\sigma}^\dagger c_{j,\sigma},  
\end{align}
which differs from the Hamiltonian Eq.~(\ref{p1_H11}) only by a background loss term.

The Hermitian part $\mathcal{H}_1$ in Eq.~(\ref{p1}) with the superlattice structure  can be constructed by loading the ultracold fermionic atoms in the lowest band of a bichromatic optical lattice \cite{PhysRevA.78.023628, Lewenstein2012, PhysRevA.100.023616, Guarrera_2007, Roati2008, Schreiber2015}, and such a Hamiltonian has been experimentally realized \cite{Schreiber2015}.  The bichromatic potential resulting from the superposition of two lattices [a main optical lattice $V_1(x)$ and a secondary weak one $V_2(x)$] can be written as \cite{PhysRevA.78.023628, Lewenstein2012, PhysRevA.100.023616}:
\begin{align}\label{p5}
	V(x) = V_1(x) +  V_2(x) = V_1\cos^2\left(k_1 x\right) + V_2\cos^2\left(k_2 x + \phi\right),  
\end{align}
where $k_i=2\pi/\lambda_i$ ($i=1,2$) are the lattice wavenumbers,  $\lambda_i$ the wavelength of the lasers which are utilized to form the optical lattices, and $V_i$ are the depth of the lattices with $V_1 \gg V_2$. The period of the superlattice potential $V(x)$ is determined by the ratio $q=k_1/k_2$. 

In the strong potential $V_1$ limit, which is much larger than the recoil energy $E_r = \hbar k_1^2/2M$ with the atomic mass $M$ \cite{PhysRevA.78.023628, Lewenstein2012, PhysRevA.100.023616}, we can only consider the lowest Bloch band. Then, the ultracold fermionic atoms in the bichromatic lattices, in the tight-binding limit, can be mapped onto the Hamiltonian $\mathcal{H}_1$ in Eq.~(\ref{p1}).

\section{Single-particle spectrum}
\label{Appendix_B}
In the absence of interactions, we consider the following single-particle non-Hermitian Hamiltonian with a superlattice potential 
\begin{align}\label{S1}
	\mathcal{H}_s =  & -t \sum_{j} \left( e^{\alpha} c_{j+1}^\dagger c_{j} + e^{-\alpha} c_{j}^\dagger c_{j+1}\right)  \nonumber \\
	&+ V_0 \sum_{j}  \cos(2 \pi j/q  + \phi) n_{j} ,  
\end{align}

The periodic modulation [i.e., the last term in Eq.~(\ref{S1})] introduces a superlattice structure, where each lattice site of the 1D chain can be now represented by two quantities: $x_m$ denoting the position of the supercell, and $\beta$ indexing the lattices inside the supercell. For a periodic system, we can transform the real-space Hamiltonian $\mathcal{H}_s$ to the momentum-space one $\mathcal{H}_s(k)$ by the following Fourier transformation
\begin{align}\label{S2}
	c_{\beta, m} = \frac{1}{\sqrt{N_{\textrm{cell}}}} \sum_k c_{\beta, k} e^{-i k x_m},  
\end{align}
where $\beta = 1, 2, \dots, q$, $N_{\textrm{cell}} = L/q$, $m= 1, 2, \dots, N_{\textrm{cell}}$, and $-\pi/q < k < \pi/q$. Thus, the momentum-space Hamiltonian is given by 

\begin{widetext}
\begin{align}\label{S3}
	\mathcal{H}_s(k) =  &-t \sum_{k} \left[e^{\alpha}\left(c_{2,k}^\dagger c_{1,k} + \dots + c_{q,k}^\dagger c_{q-1,k}  + c_{1,k}^\dagger c_{q,k} e^{i k q} \right) + e^{-\alpha} \left(c_{1,k}^\dagger c_{2,k} + \dots + c_{q-1,k}^\dagger c_{q,k} + c_{q,k}^\dagger c_{1,k} e^{-i k q} \right)\right] \nonumber \\
	& + V_0 \sum_{\beta, k}  \cos(2 \pi j/q  + \phi) n_{\beta, k} ,  
\end{align}
\end{widetext}

According to Eq.~(\ref{S3}), the single-particle spectrum is split into $q$ bands due to the superlattice potential, as shown by the gapped complex energy spectra for $q = 3$ using the periodic boundary condition in Figs.~\ref{FigNS1}(a,b). The energy spectrum is gapped for $\alpha = 0.2$ at both the $1/3$ and $2/3$ particle fillings [see Fig.~\ref{FigNS1}(c)]; while it is gapless for $\alpha = 0.4$ at the $1/3$ particle filling [see Fig.~\ref{FigNS1}(d)] due to the nonreciprocal hopping, which is different from the Hermitian case \cite{PhysRevLett.108.220401}. 

When the single-particle chain is changed from periodic boundary to the open one, in-gap modes appear for $\alpha = 0.2$ at both the $1/3$ and $2/3$ particle fillings (see Figs.~\ref{FigNS1}(e,f)]. These in-gap modes (there exist two edge modes at a specific $\alpha$) can be localized at both the left and right ends of the chain for the small nonreciprocal hopping factor $\alpha = 0.2$, as shown in Figs.~\ref{FigNS1}(g,h). While, for the strong nonreciprocal hopping, i.e., $\alpha = 0.4$, the in-gap modes are localized only at the right end due to the much larger forward hopping amplitude than the backward one [see Figs.~\ref{FigNS1}(i,j)], which is dubbed the \textit{non-Hermitian skin effect}.

\section{Neutral excitations for $U=10$}\label{Appendix_C}
In this strong interaction limit (e.g., $U=100$ in Sec.~$\textrm{\Rmnum{2}}$), the interacting system with the nonreciprocal hopping shows the anomalous boundary effect, where some in-gap states exhibit no edge excitations in the open chain. Here we present the results of the edge excitations of the neutral excitation modes for the weaker interaction $U=10$ considering the open boundary conditions, as shown in Figs.~\ref{figNS3}. As the same as the case for the strong interaction $U=100$, the gapless neutral excitation only carries a charge degree of freedom at the edges for $U=10$ [Figs.~\ref{figNS3}(a,b)]. Moreover, the in-gap modes only with $\phi < 2\pi/3$ ($\phi > 2\pi/3$) show edge excitations for $\alpha = 0.4$ ($\alpha = -0.4$), and $V_0=1.5$ [see Figs.~\ref{figNS3}(a,c)]. Therefore, the neutral excitations, calculated by only the right eigenvectors, also show the same anomalous boundary effect in the weaker interaction $U=10$ as that in the strong repulsive limit. Such an anomalous boundary effect disappears for the weak nonreciprocal hopping, i.e., $\alpha=0.2$ as shown in Fig.~\ref{figNS3}(d), or the large modulation amplitude, i.e., $V_0=4$ in Fig.~\ref{figNS3}(e). Furthermore, unlike the single-particle model, the nonreciprocal hopping here does not cause the  charge density of the ground state to accumulate near the boundaries [see Fig.~\ref{figNS3}(f)], indicating that the non-Hermitian skin effect plays no role in the particle density of the interacting system due to the combined effects of the Pauli exclusion principle and interaction. By using the biorthogonal formula, we can have the right edge excitations, as shown in Fig.~\ref{figNS3}(g,h).

%


\begin{thebibliography}{121}
\bibitem [{Not()}]{NoteMottNonHermitian}%
	\BibitemOpen
	\href@noop {} {}\bibinfo {note} {There is another phase called topological
		Mott insulator, where correlated fermions mimic gapless edge modes of bosonic
		symmetry-protected topological phases. Note that this type of phases differs
		from the non-Hermitian topological Mott insulator discussed in this
		work.}\BibitemShut {Stop}%
\end{thebibliography}
\end{document}